\documentclass[journal]{IEEEtran}
\usepackage{graphicx}
\usepackage{multirow}
\usepackage{booktabs}
\usepackage{array}
\usepackage{bm} %text bold
\usepackage{amsmath} 
\usepackage{algorithm}  
\usepackage{amssymb}
\usepackage{threeparttable}
\usepackage{CJK}
\usepackage{subfigure}

\ifCLASSINFOpdf
\else
\fi
\begin{document}

% \title{Multiscale Brain Topology Relation Reasoning for EEG-based Emotion Recognition}

\title{PGCN: Pyramidal Graph Convolutional Network for EEG Emotion Recognition}

\author{{Ming Jin, Enwei Zhu, \emph{Member, IEEE}, Changde Du, Huiguang He, \emph{Senior Member, IEEE}, Jinpeng Li*, \emph{Member, IEEE}}% <-this % stops a space
\thanks{This work was supported in part by National Natural Science Foundation of China (62106248), Zhejiang Provincial Natural Science Foundation of China (LQ20F030013), Ningbo Public Service Technology Foundation, China (202002N3181), and Medical Scientific Research Foundation of Zhejiang Province, China (2021431314).}
\thanks{Ming Jin, Enwei Zhu and Jinpeng Li are with HwaMei Hospital, University of Chinese Academy of Sciences, Ningbo, Zhejiang Province, China. They are also with Ningbo Institute of Life and Health Industry, University of Chinese Academy of Sciences, Ningbo, Zhejiang Province, China. }
\thanks{Changde Du is with Research Center for Brain-Inspired Intelligence, Institute of Automation, Chinese Academy of Sciences, Beijing, China.}
\thanks{Huiguang He is with the National Laboratory of Pattern Recognition, Institute of Automation, Chinese Academy of Sciences, Beijing, China.}
\thanks{Corresponding author: Jinpeng Li (\textit{E-mail: lijinpeng@ucas.ac.cn})}}

\markboth{Journal of \LaTeX\ Class Files,~Vol.~14, No.~8, August~2015}%
{Shell \MakeLowercase{\textit{et al.}}: Bare Demo Here of IEEEtran.cls for IEEE Journals}

\maketitle

\begin{abstract}

Emotion recognition is essential in the diagnosis and rehabilitation of various mental diseases. In the last decade, electroencephalogram (EEG)-based emotion recognition has been intensively investigated due to its prominative accuracy and reliability, and graph convolutional network (GCN) has become a mainstream model to decode emotions from EEG signals. However, the electrode relationship, especially long-range electrode dependencies across the scalp, may be underutilized by GCNs, although such relationships have been proven to be important in emotion recognition. The small receptive field makes shallow GCNs only aggregate local nodes. On the other hand, stacking too many layers leads to over-smoothing. To solve these problems, we propose the pyramidal graph convolutional network (PGCN), which aggregates features at three levels: \emph{local}, \emph{mesoscopic}, and \emph{global}. First, we construct a vanilla GCN based on the 3D topological relationships of electrodes, which is used to integrate two-order \emph{local} features; Second, we construct several mesoscopic brain regions based on priori knowledge and employ mesoscopic attention to sequentially calculate the virtual mesoscopic centers to focus on the functional connections of \emph{mesoscopic} brain regions; Finally, we fuse the node features and their 3D positions to construct a numerical relationship adjacency matrix to integrate structural and functional connections from the \emph{global} perspective. Experimental results on three public datasets indicate that PGCN enhances the relationship modelling across the scalp and achieves state-of-the-art performance in both subject-dependent and subject-independent scenarios. Meanwhile, PGCN makes an effective trade-off between enhancing network depth and receptive fields while suppressing the ensuing over-smoothing. Our codes are publicly accessible at  https://github.com/Jinminbox/PGCN.

\end{abstract}


% 需要重点参考的文献：
% 1、情感脑机接口研究综述-吕宝粮

\IEEEpeerreviewmaketitle

\section{Introduction}

% 第一段（为什么要进行情绪识别）：
\IEEEPARstart{E}{motion} recognition is an important module in human-computer interaction \cite{cowie2001emotion}, mental disease diagnosis and rehabilitation \cite{zotev2020emotion, carpenter2018cognitive}, transportation \cite{wu2020detecting} and security \cite{katsis2008toward}.

% 第二段（目前的情绪识别手段）：当前的情绪识别手段大多分为两种类型，一种是通过语言，动作和面部表情等方式，这种方式获取情绪的成本低，但是精度欠佳；相对于行为信号，生理信号所包含的情绪不易被伪装，具有更高的可靠性和准确性。常见的可以用于情绪识别的生理信号包括ECG、EMG和EEG等。
Existing works on emotion recognition fall into two categories: The first category uses low-cost, easily accessible behavioral signals such as speech \cite{gu2018deep}, gesture \cite{noroozi2018survey} and facial expression \cite{zeng2018facial}. The second category uses physiological signal, which has higher reliability and accuracy due to the robustness to artifacts and concealment \cite{nam2020disguised}. The physiological signals include EEG \cite{song2018eeg}, electrocardiogram (ECG), electromyogram (EMG) and galvanic skin response.

\begin{figure*}[h]
  \centering
    \includegraphics[width=0.95\textwidth]{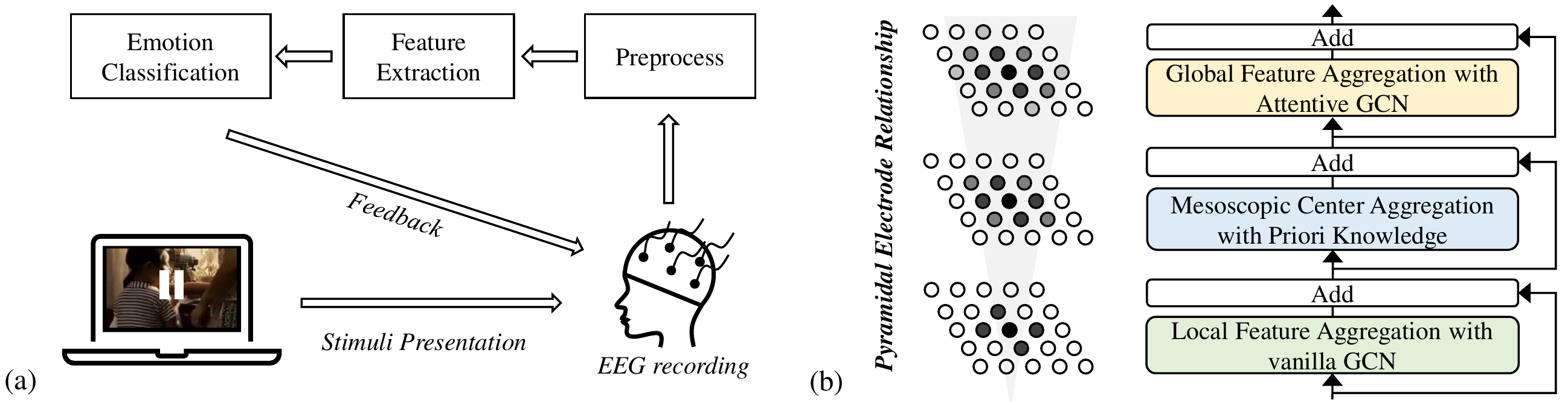}
  \caption{The EEG emotion recognition paradigm and the conceptual design of PGCN. (a) The basic flowchart of EEG emotion recognition, which consists of visual-audio stimuli presentation (to evoke corresponding emotions), EEG signal acquisition, pre-processing, feature extraction, emotion classification and (sometimes) feedback to the subject. (b) There are three main components for feature aggregation in the PGCN. The vanilla GCN extracts local bias information from neighboring nodes, the virtual mesoscopic center aggregates information within brain regions constructed with priori knowledge guidance, and the attentional GCN further fuses structural and functional connectivity at a global level.}
  \label{figure1}
\end{figure*}

% 在众多的生理信号中，EEG因其直接采集人类的情绪相关的脑部电流信号而获得了长效的发展。然而，EEG信号获取的是大脑表层的富含噪声的信号，在采集过程中非常容易受到人体自身的其他生理信号以及外界环境的干扰，需要对脑电信号进行预处理和特征提取使得其中富含的情绪信号暴露出来。预处理包括去除眼电伪迹、肌电伪迹、心电伪迹、皮肤电活动干扰、工频干扰等噪声与干扰。针对性的特征提取有助于将EEG信号中的情绪相关的信息暴露出来，有助于实现更准确的情绪识别。
EEG has achieved longevity among physiological signals due to its direct acquisition of emotion-related brain signals. However, the EEG acquisition process is very susceptible to interference from other physiological signals of the human body and environmental noise. Therefore, preprocessing and feature extraction are required to expose the emotional information contained in the EEG signal \cite{robbins2020sensitive}, Figure \ref{figure1} (a) illustrates the flowchart of EEG emotion recognition. Preprocessing includes removing noise and interference such as electrooculogram artifacts, EMG artifacts, ECG artifacts, galvanic skin, and power frequency interference. Feature extraction helps expose emotion-related information in EEG signals and helps to achieve more accurate emotion recognition.

% 第三段（分析现有基于GCN的EEG的情绪识别方法的不足）：在众多基于脑电图的情绪识别方法中，借助大脑表层不同节点的拓扑结构构建图卷积模型深受喜爱。列举基于GCN的情绪识别的诸多方法。
% The method of constructing graph convolution models with the help of the topology of different nodes on the brain scalp is prevalent due to the possibility of exploiting the connection relationships between brain electrodes. As the pioneering work of GCN in EEG emotion recognition, DGCNN \cite{song2018eeg} dynamically updates the connections between EEG channels through gradient backpropagation to obtain more discriminative EEG features. The effective combination of Graph Convolutional Broad Network and broad learning system ensures that GCB-net \cite{zhang2019gcb} can search features in broad spaces while exploring the deeper-level EEG information. RGNN \cite{zhong2020eeg} proposed two regularizers, node-wise domain adversarial training, and emotion-aware distribution learning, to handle cross-subject EEG variations better. In order to deal with the individual differences and the dynamic, uncertain relationships among different EEG regions, a variational instance-adaptive graph method (V-IAG) \cite{song2021variational} was proposed and achieved good results.

The method of graph convolution models with the help of the topology of different nodes on the brain scalp is prevalent due to the possibility of exploiting the connections between brain electrodes. As the pioneering work of GCN in EEG emotion recognition, DGCNN \cite{song2018eeg} dynamically updates the connections between EEG channels through gradient backpropagation to obtain more discriminative EEG features. GCB-net \cite{zhang2019gcb} combines GCN with broad learning system to search features in broad spaces while exploring the deeper-level EEG information. RGNN \cite{zhong2020eeg} proposed two regularizers to handle cross-subject EEG variations better. V-IAG \cite{song2021variational} was proposed to deal with the individual differences and the dynamic, uncertain relationships among EEG regions.

% 第四段（不足-->改良）：尽管之前的工作尝试从各种角度去优化面向EEG情绪识别的GCN，但是依旧面临着很多问题。首先是现有的图关系构建方法较为粗糙：最常用的图关系大多通过部署在头皮上的电极之间的相邻关系来构建，缺乏探索更深入的结构性连接和功能性连接构造。第二，人类情绪的脑部作用机理的相关研究由来已久，且已经收获了丰富的研究成果，然而据我们所知，尚没有将脑科学的研究成果应用到神经网络架构中用于提升情绪识别的效果的先例。最重要的，由于GCN经过多层堆叠后容易导致过平滑，如今主流的EEG情绪识别工作中的GCN往往为浅层网络（2-3层），但是由于人脑具备长距离连接，浅层网络容易丢失人脑的长距离依赖关系，构建一个有效的local-meso-global的情绪识别框架十分必要。
Although previous work has attempted to adopt GCN to EEG emotion recognition, there are still boundaries to explore.
\begin{itemize}
\item[(1)]The existing methods for graph construction are relatively rough: most methods for constructing adjacency matrix are based on whether the electrodes deployed on the scalp are adjacent to each other, with limited exploration of more in-depth structural and functional connectivity. 
\item[(2)]The emotion activation patterns on the brain scalp have been investigated for a long time. Nevertheless, to our knowledge, no precedent work has integrated these findings (e.g., functional clusters of electrodes) to the network architecture design in emotion recognition.
\item[(3)]Most importantly, existing GCNs in EEG emotion recognition are shallow networks (2-3 layers) because multi-layer GCN is prone to over-smoothing. However, the human brain possesses long-range connectivity, and shallow networks fail to learn such long-range dependencies.
\end{itemize}

% 第五段（PGCN方法的详细描述）：为了解决上述问题，我们提出了一种基于图卷积的从局部到介观再到全局的金字塔型网络PGCN，并将部分神经科学的研究融合其中。如图1所示，所提出的PGCN包含三个主要部分。第一部分关注不同节点的小世界特性：基于电极的3D空间距离构建邻接矩阵，并使用两层图卷积网络融合临近节点间的带有强烈区域特异性的结构性关联；第二部分关注特定脑区域节点之间的功能性联系：根据脑科学研究的部分成果划定了不同的介观脑区域，计算每个区域内节点间的注意力相关系数并生成虚拟的介观中心节点。第三部分关注全脑的节点之间可能的长距离依赖:我们使用注意力计算出融合节点间数值关系和位置关系的邻接矩阵，并使用图卷积网络融合全脑尺度的节点关系。最后，将不同尺度的特征与原始特征进行融合，并使用两层全连接网络进行最终的情绪识别。
To address these issues, we propose a graph-based pyramidal network that progressively extends the perceptual field from local to mesoscopic and global and incorporates the priori knowledge from neuroscience research. As shown in Figure \ref{figure1} (b), PGCN contains three main components. The first part focuses on strong local connections between different nodes. It constructs adjacency matrices based on 3D spatial distances of different electrodes and employs a two-layer GCN to fuse structural associations between adjacent nodes with strong region specificity. The second part focuses on the functional connections between nodes in specific brain regions. Different mesoscopic brain regions are delineated based on the brain research prior \cite{he2007small, hagmann2008mapping}. The attention correlation coefficients between nodes within each region are calculated, and virtual mesoscopic nodes are generated. The third part focuses on possible long-distance dependencies between different nodes of the whole brain. We use attention to compute global adjacency matrices that fuse numerical and positional relationships between nodes and use GCN to fuse node relationships at the whole-brain scale. Finally, the features at different scales are fused with the original features, and a 3-layer fully connected network is used for final emotion recognition.

% 第六段（该方法与之前方法的对比）：与之前的工作大多仅使用电极间的2D关系用于图卷积网络的构建相比，PGCN能够巧妙的将节点之间的绝对位置关系、相对位置关系和数值关系融合到网络中，并借鉴了情绪相关的神经科学先验研究，构建出对应的虚拟的脑区介观中心，最终构建了一个从局部，介观到全局的金字塔型图卷积网络提取不同尺度的EEG特征。我们在3个数据集上评估了网络的性能，实验结果证明所提出的PGCN取得了最佳的实验结果。此外，可视化的结果也进一步证明了网络的有效性。我们的工作的主要贡献整理如下：（1）我们构建了一个适用于EEG情绪识别任务的金字塔型图卷积网络，用以提取不同尺度的节点特征；（2）我们在情绪相关的神经科学的先验研究基础上，构建了有效的介观虚拟中心，用于增强部分脑区域在PGCN中的作用；（3）构建的网络能够提取不同节点间的结构性联系和功能性联系，有效的挖掘蕴含在节点的空间关系和节点特征中的信息，并以此来提高情绪识别的效果。

Compared with previous works that mostly use only 2D electrode relations for GCN construction, the PGCN fuses absolute positional, relative positional, and numerical relationships between nodes into the network while drawing on a priori studies in emotion-related neuroscience to construct virtual mesoscopic centers of brain regions. To achieve this goal, we construct the PGCN that aggregates EEG features at different scales from local, mesoscopic to global. The main contributions of this work is threefold: 
\begin{itemize}
\item[(1)] We exploit the \textbf{P}yramidal \textbf{G}raph \textbf{C}onvolutional \textbf{N}etwork that disposes EEG electrode features at different scales. PGCN effectively excavates the information embedded in the node's structural and functional connections, thus improving the network's effectiveness.
\item[(2)] Based on priori knowledge in emotion-related neuroscience, we design different mesoscopic regions and calculate their virtual centers to distinguish the role of different brain regions in emotion recognition tasks.
\item[(3)] We evaluate the network's performance on three open datasets. Experimental results demonstrate that PGCN achieves state-of-the-art performance. In addition, the visualization results further demonstrate the effectiveness of the proposed method. The code for the paper will be open sourced after publication.
\end{itemize}

% 第八段（论文接下来的组织）：论文的后续组织如下：第二节简单回顾近期的相关工作；第三节对于所提出的PGCN进行详细的解读；PGCN在3个情绪识别数据集上的实验结果在第四节；第五节尝试对PGCN进行更深入的分析、第六节阐述我们的结论。

The rest of this paper is organized as follows: Section \ref{Related_work} gives a brief review of related work; Section \ref{Methodology} provides a detailed interpretation of the proposed PGCN; the experimental results of the PGCN on three emotion recognition datasets are presented in Section \ref{Experiments}; Section \ref{Discussion} attempts a more in-depth analysis of the PGCN and their visualization; Section \ref{Conclusion} elaborates our conclusions.

\section{Related Work}\label{Related_work}

% 相关工作的组织：情绪识别任务，需要缩减；图卷积网络；自注意力机制；情绪与人脑；

\subsection{EEG Emotion Recognition}

% （简单介绍预处理）
Since the collected EEG raw data contain artifacts and noise, it is challenging to directly apply EEG raw data for emotion recognition. Pre-processing operations such as filtering, baseline correction, and re-referencing can effectively suppress or eliminate the artifacts and noises. Some work has reported using pre-processed EEG signals for end-to-end emotion recognition. TSception \cite{ding2020tsception} uses temporal and spatial convolutional layers to extract the time-frequency characteristics and the difference between the left and right hemispheres for end-to-end emotion recognition. EEGNet \cite{lawhern2018eegnet} uses deep separable convolutions to extract Spatio-temporal patterns and is prominent in various BCI tasks. 

% （简单介绍特征提取）
Although the preprocessed EEG data can be directly used for emotion recognition, it is more effective to further extract emotion-related EEG features. After spectral analysis \cite{cohen2014analyzing} and frequency band interception, frequency-domain feature extraction can obtain rhythmic information of neural activity in the brain. Commonly used frequency domain features include power spectral density (PSD) features \cite{kemp2004gender}, differential entropy (DE) features \cite{duan2013differential}, differential asymmetry (DASM), and Rational Asymmetry (RASM), etc.

% （介绍基础的脑电情绪识别方法)
After extracting the EEG features, different models are designed for emotion recognition. Traditional machine learning methods such as support vector machine (SVM) \cite{suykens1999least}, and clustering \cite{liang2019unsupervised} have been shown to perform emotion recognition. Deep learning has become mainstream for processing EEG features with its rapid development in recent years. Zheng \emph{et al.} \cite{zheng2015investigating} constructed a deep belief network (DBN) to explore the role of different frequency bands and channels. Yang \emph{et al.} \cite{yang2017eeg} proposes a hierarchical network structure with sub-network nodes to discriminate human emotions. Since human emotions have continuity in time, recurrent neural network (RNN) or long short-term memory can effectively utilize the temporal correlation of EEG \cite{alhagry2017emotion}, ACRNN \cite{tao2020eeg} employed CNN to extract the spatial information and applied RNN with extended self-attention to explore the temporal information of EEG feature, Zhang \emph{et al.} \cite{zhang2019making} introduced CNN and RNN to explore the preserved spatial and temporal information in either a cascade or a parallel manner.

\subsection{Graph Convolutional Networks}

Thanks to the introduction of practical structural information, GCN is considered an effective method for processing non-Euclidean data and has achieved great success in the field of social networks \cite{kipf2016semi}, knowledge graphs \cite{hamaguchi2017knowledge} and traffic prediction \cite{zhao2019t}, and so on.

An undirected graph can be expressed as $\mathcal{G}=(\mathcal{V}, \mathcal{E})$, in which $\mathcal{E}$ represents the set of edges that connect different nodes in the set of $\mathcal{V}$. By connecting all edge $\mathcal{E}$, an adjacency matrix $\mathbf{A} \in \mathbb{R}^{n \times n}$ that characterizes the graph relationship can be constructed, at the same time, by aggregating the feature $\mathbf{x} \in \mathbb{R}^{d}$ of each node, a feature matrix $\mathbf{X} \in \mathbb{R}^{n \times d}$ characterizing the features of the nodes on the graph can be constructed, where $n$ denotes the number of nodes and $d$ is the dimension of input features. 

The simplified GCN proposed by Kipf \emph{et al.} \cite{kipf2016semi} effectively simplifies the original complex spectral GCN method \cite{defferrard2016convolutional}, and the simplified GCN can be expressed as
\begin{equation}
    \mathbf{Z}=\sigma \left(\tilde{\mathbf{D}}^{-\frac{1}{2}} \tilde{\mathbf{A}} \tilde{\mathbf{D}}^{-\frac{1}{2}} \mathbf{X} \mathbf{\Theta} \right),
\end{equation}
where $\tilde{\mathbf{A}} = \mathbf{A} + \mathbf{I}$ and $\tilde{\mathbf{D}}_{ii} = \sum_{j} \tilde{\mathbf{A}}_{i j}$, $\mathbf{I} \in \mathbb{R}^{n \times n}$ is the n-dimensional degree matrix, $\mathbf{\Theta}$ is a trainable weight matrix.

\subsection{GCN in EEG emotion recognition}

% 由于图卷积可以有效的处理非网格形式的脑电数据，基于图卷积的情绪识别也得到了迅速的发展。DGCNN\cite{song2018eeg}通过梯度反向传播动态更新表征节点之间关系的邻接矩阵，获得了更准确的节点间关系和更好的情绪识别效果。RGNN\cite{zhong2020eeg}使用节点式领域对抗训练和情绪感知分布学习两个正则器，有效优化了跨被试的情绪识别效果。受大脑认知过程的神经学知识的启发，LGG-net \cite{ding2021lggnet}提出了局部和全局图形过滤层来学习大脑不同功能区域内和之间的大脑活动，以模拟人类大脑认知过程中的复杂关系,从而实现了最佳的情绪识别效果。为了处理个体差异和不同EEG区域之间的动态不确定关系，（V-IAG）\cite{song2021variational}提出了变异实例自适应图方法并取得了良好的效果。
Since the EEG feature is a specific structured non-Euclidean data, GCN-based EEG sentiment recognition has been developed rapidly. DGCNN \cite{song2018eeg} dynamically updates the adjacency matrix characterizing the relationship between nodes by gradient backpropagation to obtain more accurate inter-node relationships and better emotion recognition. RGNN\cite{ zhong2020eeg} used two regularizers, node-based domain adversarial training, and emotion-aware distribution learning, to optimize the cross-subject emotion recognition effect. Inspired by the neurological knowledge of brain cognitive processes, LGG-net \cite{ding2021lggnet} proposed local and global graphical filtering layers to learn brain activities within and between different brain functional regions to simulate the complex relationships in human brain cognitive processes, thus achieving the best emotion recognition effect. LR-GCN \cite{jin2021eeg} employed self-attention forward updating Laplacian matrices and gradient backpropagation updating adjacency matrices to construct learnable brain electrode relationships jointly. To deal with individual differences and dynamic, uncertain relationships between different EEG regions, (V-IAG)\cite{song2021variational} proposed the variational instance adaptive graph method and achieved good results.

\subsection{Emotion and Human Brain}

% 人类对于脑网络的研究由来已久，对于脑网络的深入探究能够促进对于人类情绪的理解。Sporns等人提出可以从微尺度（microscale）、中间尺度（mesoscale）和大尺度（macroscale）上从事人脑连接组相关的研究，分别对应神经元、神经元集群和大脑脑区三种研究范围。但是，由于人类的神经元数量巨大，神经元级别的研究尚不现实，绝大部分的工作专注于神经元集群及更大的尺度的研究。
Brain networks have been studied for a long time, with a view to contributing to the understanding of human emotions. Sporns \emph{et al.} \cite{sporns2005human} proposed that studies related to the human brain's connectome can be conducted at the microscale, mesoscale, and macroscale, corresponding to neurons, neuronal clusters, and brain regions, respectively. However, due to the vast number of neurons in human brain, neuron-level studies are not yet practical, and most of the work focuses on neuronal clusters and larger scales.

% 人脑具有显著的小世界特性，表现为相邻的神经元更容易进行频繁的信息交换。He等人{62}采用124个被试的结构像数据首次构建了一个由54个节点构成的脑网络，并观察到网络中的小世界特性。后续的许多工作都证实了不同的节点划分方式下的小世界特性，且节点度的分布都服从指数截尾的幂律分布{63，64}。Hagmann等人{65}使用diffusion spectrum imaging, DSI技术构建了5个被试的998个脑区和66个分区的加权脑结构网络，通过研究发现脑网络还可以划分为6个更大的模块，且每个模块具有对应的核心节点，这些核心节点主要分布在额叶、颞叶和枕叶。
The human brain has significant small-world properties, as demonstrated by the fact that neighboring neurons are more likely to exchange information frequently. He \emph{et al.} \cite{he2007small} used structural image data to construct a 54-node brain network for the first time and observed small-world properties. Many subsequent works have confirmed the small-world properties with different node partitioning methods, and the distribution of node degrees obeyed the power-law distribution with exponential truncation tails \cite{gong2009mapping}. Hagmann \emph{et al.} \cite{hagmann2008mapping} used the diffusion spectrum imaging technique to construct a weighted brain structural network and found that the brain network could be divided into six modules, and each module had corresponding core nodes, which were mainly distributed in the frontal, temporal and occipital lobes.

% 为了进一步的提升神经元之间的通信效率{66，67}，大脑也被发现具有显著的“长距离连接”。虽然长距离连接在能量和体积上所需要的代价比短距离连接更高，但是却大大减少了数个间接的短距离连接之间的布线成本，从而提供了更快，更直接和更小噪音的信息传递{68}。
To further enhance communication efficiency between neurons \cite{bullmore2012economy}, the brain has been found to have significant "long-range connections." Although long-distance connections are more costly in energy and volume than short-distance connections, they significantly reduce the cost of wiring between several indirect short-distance connections, thus providing faster, more direct, and less noisy information transport \cite{buzsaki2004interneuron}.

\begin{figure*}[h]
  \centering
  \includegraphics[width=1.9\columnwidth]{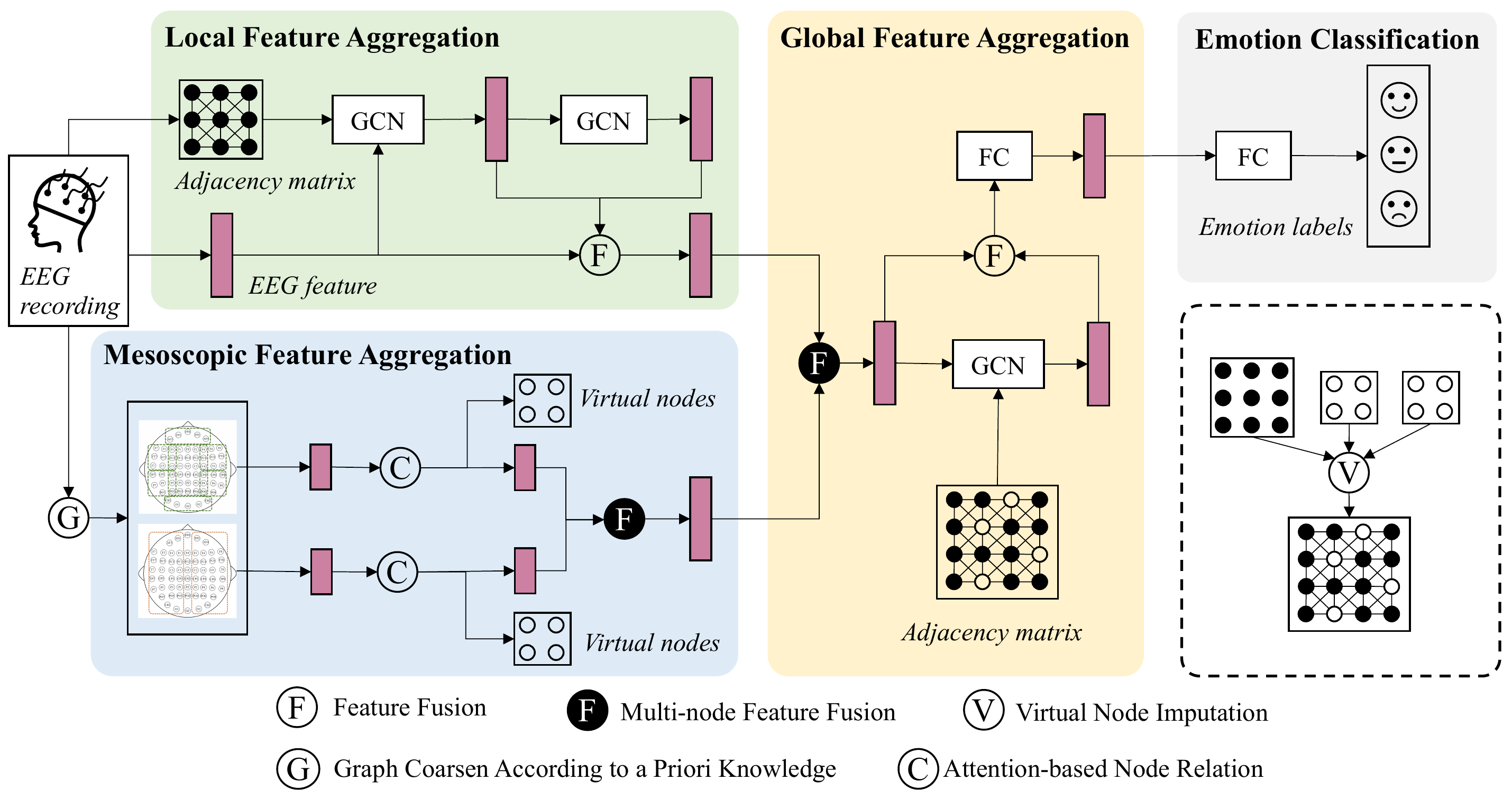}
  \caption{The flowchart of the proposed PGCN. To excavate the electrode relationships, PGCN aggregates multiscale information, i.e., local, mesoscopic and global features to conduct the emotion classification task. GCNs are used to model the relationships.}
  \label{figure2}
\end{figure*}

\section{Methodology}\label{Methodology}

\subsection{Overview}

% PGCN通过建立一个金字塔图卷积网络，逐层融合不同尺度的EEG特征，增强了基于EEG的情绪识别能力。该模型的整体架构如图2所示，可分为以下步骤。(1)为了融合具有强局部相关性的EEG特征，我们根据电极之间的空间距离构建一个稀疏的结构连接矩阵，并引入图卷积网络来聚合局部特征。(2)由于不同脑区对于情绪的响应程度不同，为了更好地区分不同脑区对情绪的相关性，我们根据先验研究构建了中观尺度的脑区，并计算每个脑区的虚拟介观中心来表征介观特征。(3)为了避免不同节点之间的长距离连接丢失，我们将原始节点与虚拟中观节点进行融合，借助注意力机制构建全局图连接网络，并利用图卷积融合全局特征。(4)最后，将每个尺度的融合特征送入由3层全连接网络组成的情感识别任务网络，输出最终的情感标签。
PGCN builds a pyramidal network that fuses EEG features at different scales layer by layer. The overall architecture of the model is shown in Figure \ref{figure2}, which can be divided into the following steps. (1) Considering the frequent local
connections of brain networks, we construct a sparse structural adjacency matrix based on the spatial distance between electrodes and introduce a GCN to aggregate local features. (2) To better distinguish the relevance of different brain regions to emotions, we constructed mesoscopic-scale brain regions based on a priori studies and calculated virtual mesoscopic centers for each brain region to characterize the mesoscopic features. (3) To balance the importance and economics of long-distance connections, we fuse the original nodes with virtual mesoscopic nodes, construct a sparse global graph connectivity network with the help of an attention mechanism, and aggregate global features with graph convolution. (4) Finally, the fused features are fed into 3-layer fully connected network for final emotion recognition tasks.

\subsection{Local Feature Aggregation}

% 局域特征聚集专注于人脑的小世界属性，构建有效的信息传递机制融合邻近节点的脑电信息。为了专注于短距离邻居特征，我们首先构建了基于电极的空间相对位置的邻接矩阵，并使其度服从指数截尾的幂分布\cite{iturria2008studying, gong2009mapping}。此外，考虑到人脑神经元布线的高效原则{bullmore2012economy, achard2007efficiency}，我们保持了邻接矩阵具备足够的稀疏性。构建出合适的邻接矩阵后，我们进行了两次图卷积操作聚合邻接节点的特征。之后，我们将原始特征与图卷积之后的特征进行融合，用以抑制信息传递后的过平滑{li2021deepgcns}。
Local feature aggregation focuses on the frequent local connections of the human brain and constructs an effective information transport mechanism to fuse the EEG information of neighboring nodes. To focus on short-range neighboring features, we first constructed sparse graph relations reasoning based on the relative spatial positions of electrodes and made their degrees obey exponentially truncated power distributions \cite{gong2009mapping} and kept the adjacency matrix with sufficient sparsity \cite{zhong2020eeg}. After that, we constructed a two-layer GCN to aggregate the features of the adjacency nodes. Finally, we fused the original features with the features after graph convolution to suppress over-smoothing \cite{li2021deepgcns}.

\subsubsection{Sparse Graph Relation Reasoning}

% 图卷积网络非常依赖构建准确的节点关系矩阵，并计算出对应的拉普拉斯矩阵。在脑电情绪识别的任务中，常用的邻接矩阵的构建方式是基于电极的相邻关系构建：
GCN relies on constructing an accurate node relation matrix and calculating the corresponding Laplacian matrix. In the task of EEG emotion recognition, a common way to construct the adjacency matrix of different electrodes {$i$} and {$j$} is:

\begin{equation}
\mathbf{A}_{ij}=\left\{
\begin{aligned}
1 \quad  \text{if} \quad {j} \in \mathcal{N}_{i} \\
0 \quad  \text{if} \quad {j}  \notin \mathcal{N}_{i}
\end{aligned}
\right.
,
\end{equation}
where $\mathcal{N}_{i}$ represents the 2D spatial neighbor of electrode $i$. 

% 但是，这种构建方式面临着一些问题：首先，市面上所有的脑电极帽的电极都无法做到均匀分布，仅使用简单的0/1无法准确的衡量电极之间的关系；其次，构建的邻接矩阵会导致过稀疏问题，导致在模型的优化过程中丢失部分关键连接；最后，由于脑电具有非常大的跨被试甚至跨session差异，构建固定不变的邻接矩阵会导致模型的欠优化问题。
However, the method faces some problems: (1) The electrodes of all commercially available brain electrode caps cannot be evenly distributed, and the relationship between electrodes cannot be accurately portrayed by simple numbers 0 and 1. (2) The constructed adjacency matrix will lead to over-sparsity, resulting in the loss of some critical connections in the optimization process. (3) Since EEG has very large cross-subject or even cross-session differences, the construction of a fixed adjacency matrix will lead to under-optimization.

% 受到hfd和jhdfie的工作的启发，为了更好的描述不同通道之间的空间关系，我们构建了基于通道之间距离平方反比的初始化邻接矩阵：
Inspired by the work of Salvador \emph{et al.} \cite{salvador2005neurophysiological} and Zhong \emph{et al.} \cite{zhong2020eeg}, to better describe the connections between different nodes, we constructed an initial adjacency matrix based on the inverse square of the spatial distance between different nodes:

\begin{equation}
%   \mathbf{A}_{i j}=\frac{\delta}{d_{i j}^{2}} \quad \text{and} \quad 0.1 \le \mathbf{A}_{i j} \le 1,
%   \begin{equation}
\mathbf{A}_{ij}=\left\{
\begin{aligned}
& 1 \quad &\text{if} & \quad \mathbf{A}_{i j} \ge 1\\
& \frac{\delta}{d_{i j}^{2}} \quad &\text{if} & \quad 0.1 \le \mathbf{A}_{i j} \le 1 \\
& 0.1 \quad &\text{if} & \quad \mathbf{A}_{i j} \le 0.1
\end{aligned}
\right.
,
\end{equation}
% \end{equation}
where ${d_{i j}}$ is 3D distance between node ${i}$ and ${j}$, and $\delta$ is the sparsity factor. In the experiment, we found that the best result is obtained when $\delta$ is set to 9. We clip $\textbf{A}_{i j}$ greater than 0.1 to maintain the sparsity; and clip $\textbf{A}_{i j}$ less than 1 to reduce the weights of the self-loop and extremely close neighbors.

Refering the work of Jin \emph{et al.} \cite{jin2021eeg} in LR-GCN, we set $\mathbf{A}$ as a learnable matrix and update it through gradient backpropagation.

\subsubsection{GCN for Local Representation Aggregation}

% 构建完合适的邻接矩阵A后，我们计算出对应的laplacian矩阵L并使用GCN执行邻居节点间的信息传递。
After constructing the appropriate adjacency matrix $\mathbf{A}$, we compute the corresponding laplacian matrix $\mathbf{\hat L}$ and perform information transport between neighbor nodes with GCN:
\begin{equation}
  \mathbf{H}^{(l+1)}=\sigma\left(\mathbf{\hat L} \mathbf{H}^{(l)} \mathbf{W}^{(l)}\right),
\end{equation}
where $\mathbf{H}^{(l)}$ and $\mathbf{H}^{(l+1)}$ are the input and output node representations at layer $l$, respectively; the initial input representations $\mathbf{H}^{(0)}$ are the original input features $\mathbf{X}$. $\mathbf{W}^{(l)}$ is a learnable weight matrix and $\sigma$ is an activation function.

% 为了防止过平滑对于网络的影响，我们仅设计了两层GCN网络。此外，为了进一步的减少过平滑的影响，我们在网络中引入了跨层连接，最终的局域特征聚集的输出特征为：
In order to prevent the GCN from over-smoothing, we designed only a two-layer local GCN network and introduced cross-layer connections; the final output of the local feature aggregation is:

\begin{equation}
  \mathbf{X}^{local}=\text{concat} \left(\mathbf{X}, \mathbf{H}^{(1)}, \mathbf{H}^{(2)}\right)
\end{equation}

\subsection{Mesoscopic Feature Aggregation}

Different areas of the human cerebral cortex are highly connected and centralized, and some work has been reported dividing the cortex into several emotion-related brain regions \cite{hagmann2008mapping, bruder2017right}. With reference to priori knowledge of current brain science research, we constructed different mesoscopic brain regions and then calculated the location and feature of the virtual mesoscopic center in each region. Since the features in the region only converge to the virtual mesoscopic center, it can effectively avoid the over-smoothing while increasing the perceptual field.

\subsubsection{Graph Coarsen for Mesoscopic Regions}
\begin{figure}[ht!]
	\centering
	\mbox{
		\subfigure[7 mesoscopic regions ]{\includegraphics[width=0.44\linewidth]{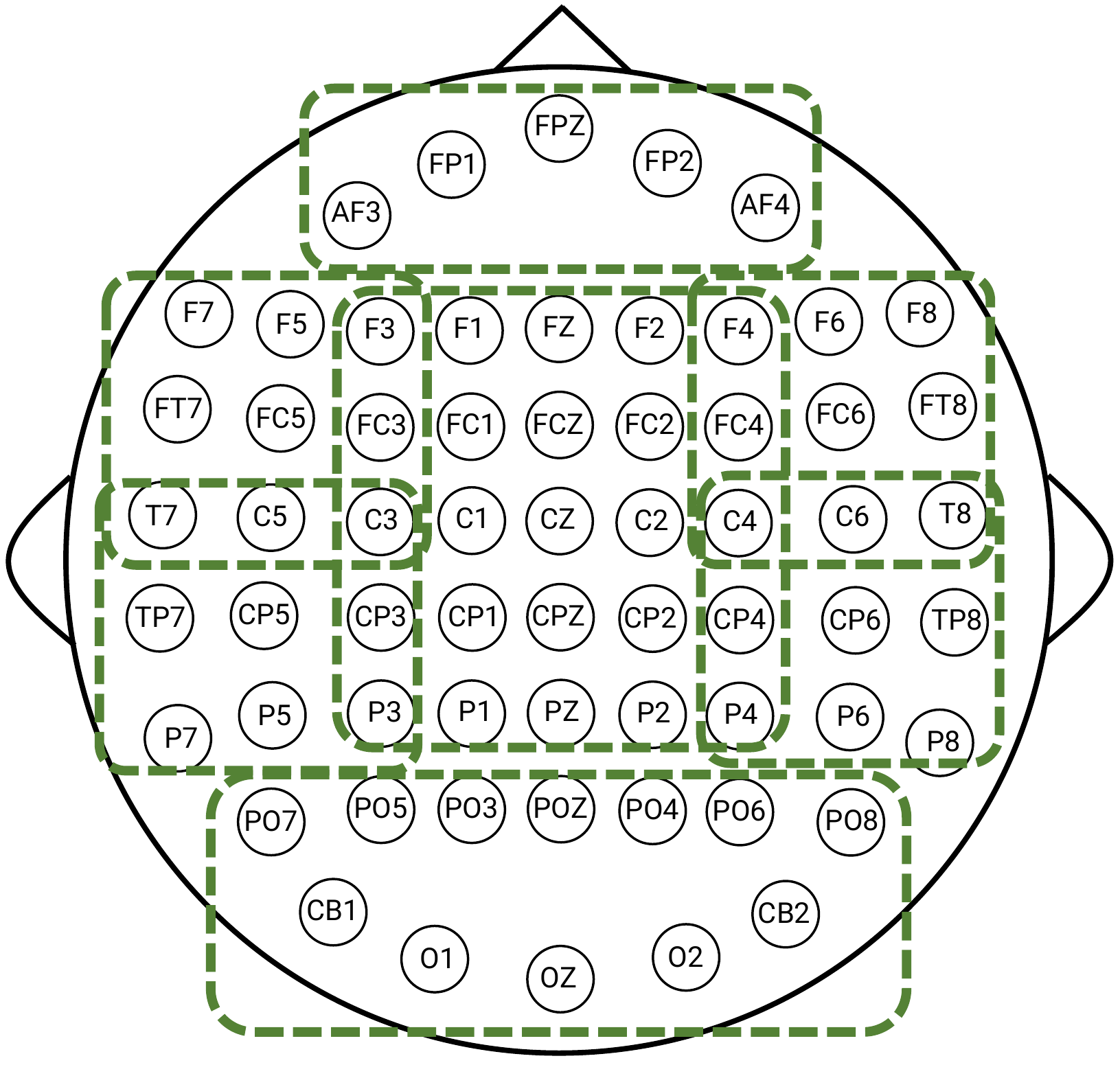}\label{ini_chord}}
      \subfigure[2 mesoscopic regions ]{\includegraphics[width=0.44\linewidth]{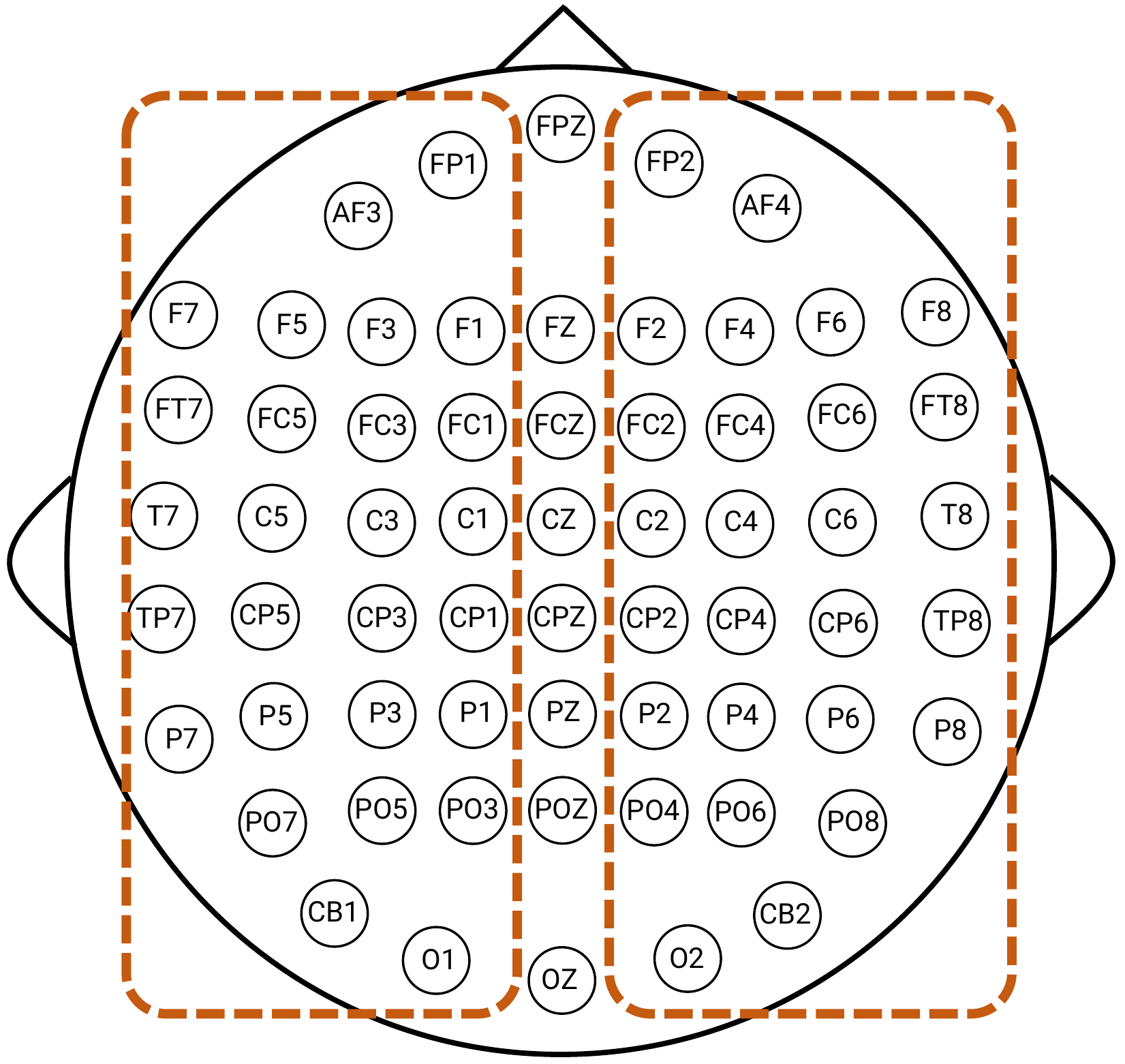}\label{att_chord}}
	    }
	\caption{Brain region division based on priori knowledge.}
	\label{figure3}
\end{figure}

% 为了获取到介观尺度的特征聚合，我们设计了两种介观尺度的节点区域划分方式，并在各自划定的脑区域内计算出虚拟的区域中心的特征和位置。
To obtain the mesoscopic feature aggregation, we designed two different mesoscopic divisions with different sensory fields, as shown in Figure \ref{figure3}.

Figure \ref{figure3}(a) shows the first partition with reference to the anatomy of the cerebral cortex. The brain's cortex is generally divided into four lobes, frontal, parietal, temporal, and occipital \cite{jawabri2019physiology}, each lobe being responsible for a different task; for example, the occipital lobe is associated with visual processing and interpretation. Firstly, we divided the electrodes into five regions; secondly, as electrodes located in the temporal lobe (FT7, T7, TP7, FT8, T8, TP8) have been reported to be important for emotion recognition, we performed a more detailed secondary division of the electrodes in the temporal lobe \cite{zheng2015investigating, zheng2017identifying}; finally, inspired by the work of Hagmann \emph{et al.} \cite{hagmann2008mapping}, we adjusted the division of the regions to meet the needs of both structural and functional connectivity.

% 人类的大脑的两个半球通过胼胝体连接构成，人类的所有活动都是两个半球的信息交互下实现。但是，两个半脑在功能上具有非常明显的差别，且有各自负责的功能。另一种有效的介观区域划分方式基于大脑的左右半球进行划分，详细的划分方式如图3(b)所示。
Figure \ref{figure3}(b) shows the other valid way to devide the mesoscopic regions based on the brain's two hemispheres. The corpus callosum connects the two hemispheres of the brain, and all human activities are realized through their information interaction. However, the functions of the two halves of the brain differ significantly.

\subsubsection{Mesoscopic Node Relation}

After constructing the mesoscopic brain regions, we focused on functional connectivity within the mesoscopic regions through a self-attention. The input is the set of features within each mesoscopic region, $\mathbf{h}=\left\{\vec{h}_{1}, \vec{h}_{2}, \ldots, \vec{h}_{N}\right\}$, $\mathbf{h} \in \mathbb{R}^{N\times F}$, where $N$ is the number of nodes in the mesoscopic region, and $F$ is the number of features in each node. The attention-based connectivity matrix $\mathbf{e}$ can be expressed as:

\begin{equation}
  \mathbf{e} = \text{LeakyReLU}((\mathbf{hW})(\mathbf{hW})^T),
  \label{equ_att}
\end{equation}
where $\mathbf{W}$ is a learnable weight matrix and $\cdot^T$ is transposition.

\subsubsection{Virtual Mesoscopic Center}
% 计算出介观区域的attention-based connectivity matrix后，依此计算虚拟介观中心的节点特征和位置。首先，我们对$e$执行按行相加，计算出介观区域内每一个节点的权重系数。获取到区域内的节点权重后，我们使用权重分别计算虚拟介观区域中心的位置和特征：
After calculating the attention-based connectivity matrix of the mesoscopic region, we begin to construct the virtual mesoscopic centers. We compute the weight coefficient $\mathbf{\Lambda} \in \mathbb{R}^{1 \times N}$ in each mesoscopic region by performing a row-wise summation for $\mathbf{e}$, where $\mathbf{\Lambda}_{i}$ denotes the functional connection weight of each node to all other nodes. After that, we begin to calculate the features and locations of the virtual mesoscopic region center:
\begin{equation}
  \begin{aligned}
   \mathbf{p}^{locate}  & = \text{softmax}(\mathbf{\Lambda})\mathbf{P}, \\
   \mathbf{m}^{feature} & = \text{softmax}(\mathbf{\Lambda})\mathbf{h},
  \label{}
  \end{aligned}
\end{equation}
where $\mathbf{P} \in \mathbb{R}^{N \times 3}$ is the absolute position matrix of the electrodes. 

% 计算出每一个介观区域的区域中心的特征和位置后，按照不同的划分方式将特征和位置进行融合。图3(a)中融合后的特征和位置分别为 \mathbf{M}^{(1)}和 \mathbf{S}^{(1)}，图3(b)中融合后的特征和位置分别为\mathbf{M}^{(1)}和 \mathbf{S}^{(1)}。

After calculating the regional centers of each mesoscopic region, the features and locations are fused according to different divisions. The fused features and locations in Figure \ref{figure3} (a) and Figure \ref{figure3} (b) are $\mathbf{M}^{(1)} \in \mathbb{R}^{7 \times F}$ and $\mathbf{P}^{(1)} \in \mathbb{R}^{7 \times 3}$, $\mathbf{M}^{(2)} \in \mathbb{R}^{2 \times F}$ and $\mathbf{P}^{(2)} \in \mathbb{R}^{2 \times 3}$, respectively.

Finally, we implement virtual node imputation to fuse the virtual mesoscopic centers with the original electrode nodes, and obtained a fused feature and location matrix containing vital local and mesoscopic attributes.

\begin{equation}
  \begin{aligned}
  \mathbf{X}^{meso} & = \left(concat \left((\mathbf{X}^{local})^T, (\mathbf{M}^{(1)})^T, (\mathbf{M}^{(2)})^T\right)\right)^T, \\
  \mathbf{P}^{meso} &  = \left(concat \left(\mathbf{P}^T, (\mathbf{P}^{(1)})^T, (\mathbf{P}^{(2)})^T\right)\right)^T. \\
  \label{}
  \end{aligned}
\end{equation}

\subsection{Global Feature Aggregation}

% 尽管构建不同的介观分区能够有效的缓解较远距离的节点之间的信息传递的问题，但是随着节点数目的增加，构建更大范围的介观区域并计算介观中心很容易造成信息的丢失。

% 注意力机制已经被证明在获取全局的特征关系中十分有效。为了构建更远距离的节点之间的关联性，我们基于注意力机制构建了绝对位置相关的特征关联性邻接矩阵，用于表征每一个节点与所有节点之间的可能的联系，并剔除其中的弱连接关系。构建好有效的全局电极连接关系后，我们使用图卷积进行全局特征聚合。

Although mesoscopic partitions can effectively alleviate the node information cannot transport at longer distances, a more extensive range of mesoscopic regions can easily cause information loss as the number of nodes increases. 

Attention mechanisms are very effective in constructing global feature relationships \cite{velivckovic2017graph}. In order to assemble long-distance node correlations, we construct an attention-based absolute position-dependent feature correlation adjacency matrix to characterize the possible connections between each node and eliminate the weak connection relationships among them. After constructing the global electrode connectivity relationship, we use graph convolution for global feature aggregation.

\subsubsection{Global Node Relation}

% 为了更好的表征节点之间的全局关系，我们使用了基于电极的空间位置的位置编码。全部节点$P^meso$由原始的物理电极表征的节点$P$和由虚拟介观中心构成的虚拟节点$P^virtual$构成。由于虚拟节点也有对应的特征，我们对于所有的节点的特征与拓展维度的位置编码进行相加，获得位置增强的节点特征：
We established node position encoding with their spatial location to better characterize the global node relationships. The full nodes $\mathbf{P}^{meso}$ consist of the original electrode nodes $\mathbf{P}$ and the virtual mesoscopic nodes $\mathbf{P}^{virtual}$. Since the virtual nodes also have corresponding features, we sum the features of all nodes with the position encoding to obtain the position-enhanced node features as follows:

\begin{equation}
  \mathbf{X}^{enhanced} = \mathbf{X}^{meso}+embed\left(\mathbf{P}^{meso}\right).
\end{equation}

% 之后，我们使用包含6个头的多头自注意力机制分别计算出6个不同的注意力关系矩阵G，并使用可学习的权重向量将G融合成最终的注意力关系矩阵：

After that, we compute the relation matrix $\mathbf{G} \in \mathbb{R}^{6 \times N' \times N'}$ using a multi-head self-attention containing 6 heads and fuse $\mathbf{G}$ into the final attentional relation matrix $\mathbf{A}^{global} \in \mathbb{R}^{N' \times N'}$ using a learnable weight vector $\mathbf{w} \in \mathbb{R}^{1 \times 6} $:

\begin{equation}
  \mathbf{A}^{global} = \mathbf{w}\mathbf{G}.
\end{equation}

The dense attention matrix empirically exacerbate over-smoothing but we intentionally preserve the top 20\% connections in the adjacency matrix to ensure its sparsity.

\subsubsection{GCN for Global Representation Aggregation}

After obtaining the global attention-based adjacency matrix $\mathbf{A}^{global}$, we compute the corresponding laplacian matrix $\mathbf{\hat L}^{global}$ and employ it for graph convolution:

\begin{equation}
  \mathbf{O}^{(l+1)} =\sigma\left(\mathbf{\hat L}^{global} \mathbf{O}^{(l)} \mathbf{W}^{(l)}\right),
\end{equation}

where $\mathbf{O}^{(l)}$ are the input node representations and  $\mathbf{O}^{(l+1)}$ are the output node representations, the initial input representations $\mathbf{O}^{(0)}$ are the original input features $\mathbf{X}^{meso}$. $\mathbf{W}^{(l)}$ is a learnable weight matrix and $\sigma$ is activation function.

After that, we concatenate the input mesoscopic features $\mathbf{X}^{meso}$ with the global GCN output $\mathbf{O}^{(1)}$ to obtain the feature that covers the local, mesoscopic, and global perceptual fields.

\begin{equation}
  \mathbf{X}^{global}=concat \left(\mathbf{X}^{meso}, \mathbf{O}^{(1)}\right)
\end{equation}

% 之后，我们在x之后设计了三层全联接网络，输出情绪识别的最终结果。
Finally, we set $\mathbf{X}^{global}$ as the input of a three-layer fully connected emotion recognition network to obtain the final output of the subject's emotions. Details of the model implementation are in Appendix \ref{model}.

\section{Experiments}\label{Experiments}

In this section, we evaluate the effectiveness of the proposed PGCN on three well-known emotion recognition database, SEED\cite{zheng2015investigating}, SEED-IV\cite{zheng2018emotionmeter}, and SEED-V\cite{liu2021comparing}.

\subsection{Datasets and Protocol}

To comprehensively evaluate the effectiveness of the proposed PCGN, we conducted subject-dependent and subject-independent experiments on the above three datasets; pre-processing and feature extraction was carried out for subsequent objective model evaluation. In the subject-dependent experiments, the training data and testing data were both from the same subject to evaluate the effectiveness of the model for cross-temporal application to the same subject; in the subject-independent experiments, the training data and testing data were from different subjects to evaluate the effectiveness of the model for the cross-subject application. A detailed description of the dataset and protocol is in Appendix \ref{Dataset}. The experimental results for all baselines are extracted from the citations.

\subsection{Experiment on SEED}

Table \ref{table1} presents the subject-dependent emotion recognition accuracy of PGCN and all baselines on the SEED dataset, and the results of all baselines are extracted from the corresponding references. SVM is a traditional machine learning method, DBN and BiDANN-S are deep learning methods, and DGCNN, GCB-net+BLS, RGNN, and V-IAG are GCN-based methods.

%%%%%%%%%%%%%%%% table-1

\begin{table*}[]
  \centering
  \caption{Subject-dependent classification accuracy (mean/std) on the SEED dataset}
  \label{table1}
  % \begin{tabular}{lllllll}
  \renewcommand{\arraystretch}{1.5}
  \begin{threeparttable}
  \begin{tabular}{m{3cm}<{\centering}|m{2cm}<{\centering}m{2cm}<{\centering}m{2cm}<{\centering}m{2cm}<{\centering}m{2cm}<{\centering}m{2cm}<{\centering}}
  % \begin{tabular}{p{2.5cm}p{2cm}p{2cm}p{2cm}p{2cm}p{2cm}p{2cm}}
  \toprule
  \textbf{Method} &  \bm{$\delta$} \textbf{(1--3 Hz)}  &  \bm{$\theta$} \textbf{(4--7 Hz)} &  \bm{$\alpha$} \textbf{(8--13 Hz)}    & \bm{$\beta$} \textbf{(14--30 Hz)}   & \bm{$\gamma$} \textbf{(31--50 Hz)}    & \textbf{all bands }  \\ \hline 
  SVM \cite{suykens1999least}  & 60.50 / 14.14 & 60.95 / 10.20 & 66.64 / 14.41 & 80.76 / 11.56 & 79.56 / 11.38 & 83.99 / 9.72 \\
  DBN \cite{zheng2015investigating}  & 64.32 / 12.45 & 60.77 / 10.42 & 64.01 / 15.97 & 78.92 / 12.48 & 79.19 / 14.58 & 86.08 / 8.34  \\
  DGCNN \cite{song2018eeg} $^\dag$  & 74.25 / 11.42 & 71.52 / 5.99 & 74.43 / 12.16 & 83.65 / 10.17 & 85.73 / 10.64 & 90.40 / 8.49 \\
  BiDANN-S \cite{li2018bi} & 76.97 / 10.95 & 75.56 / 7.88 & 81.03 / 11.74 & 89.65 / 9.59 & 88.64 / 9.46 & 92.38 / 7.04  \\
  GCB-net+BLS \cite{zhang2019gcb} & 79.98 / 8.93 & 76.51 / 9.56 & 81.97 / 11.05 & 89.06 / 8.69 & 89.10 / 9.55 & 94.24 / 6.70 \\ 
  RGNN \cite{zhong2020eeg} $^{\dag}$ & 76.17 / 7.91 & 72.26 / 7.25 & 75.33 / 8.85 & 84.25 / 12.54 & 89.23 / 8.90 & 94.24 / 5.95 \\ 
  V-IAG \cite{song2021variational} $^\dag$   & \textbf{81.14 / 9.46} & 82.37 / 7.44 & \textbf{84.51 / 9.68} & 92.15 / 8.90 & 92.96 / 6.19 & 95.64 / 5.08  \\ \hline
  \textbf{PGCN (ours) $^\dag$}   &     79.62 /  10.53    &    \textbf{83.62} /  \textbf{6.91}    &   83.74 /  9.57    &   \textbf{92.33} /  \textbf{8.66}   &  \textbf{93.05} /  \textbf{5.78}   &  \textbf{96.93} /  \textbf{5.11}   \\
  \bottomrule
  \end{tabular}
  \begin{tablenotes}
  \footnotesize
  \item[$\dag$] calculate the average accuracy based on the results of two sessions.
  
  \end{tablenotes}
  \end{threeparttable}
  \end{table*}

Encouragingly, on the subject-dependent emotion recognition task, our PGCN achieves the best emotion recognition results on all-frequency band and theta, beta, and gamma bands and performs marginally worse than the previous best model V-IAG in delta and alpha frequencies. For the first time to our knowledge, the accuracy of emotion recognition has been increased to 96.93\%. By comparing the results of emotion recognition in different frequency bands, it can be seen that the model is better at capturing the emotional information carried in high-frequency features to achieve higher accuracy, which is in line with the findings of many previous studies \cite{zhong2020eeg, zheng2015investigating}.

% 表二展示了PGCN和所有的baselines模型在SEED上进行被试独立的情绪识别的结果，所有的baselines的结果均摘自对应的引文。其中，我们同时收集了基于监督学习的实验结果和基于迁移学习的实验结果。
Table \ref{table2} shows the results of the PGCN and all baselines models for subject-independent emotion recognition on the SEED dataset. In Table \ref{table2}, we collected the results of both supervise and transfer learning-based experiments.

% 经过对比可以发现，所提出的PGCN可以在所对比的基于监督学习的方法中达到sota，且领先去除领域自适应模块的RGNN 2.67%。由于迁移学习的方法会额外使用测试集的数据（不使用测试集的标签）用于模型训练，使得监督学习的方法显著落后于迁移学习的方法，RGNN在添加上DA模块后领先PGCN 0.71%。但是在实际在线情绪识别中，获取足够的测试集的数据用于迁移学习模型的训练几无可能。
The comparison shows that the proposed PGCN achieves the best results among the supervised learning-based methods, leading the RGNN with the domain adaptive module removed by 2.67\%. Since the transfer learning uses additional data from the testing set (without using the labels from the testing set) for model training, making the supervised learning approach significantly behind the transfer learning approach, RGNN is 0.71\% ahead of PGCN with the DA module added on. However, it is almost impossible to obtain enough data from the test set for training a transfer learning model in practical online emotion recognition.

  \begin{table}[]
  \centering
  \caption{Subject-independent classification accuracy (mean/std) on the SEED dataset}
  \label{table2}
  % \begin{tabular}{lllllll}
  \renewcommand{\arraystretch}{1.5}
  \begin{threeparttable}
  \begin{tabular}{m{2.5cm}<{\centering}|m{2.4cm}<{\centering}m{2.52cm}<{\centering}}
  % \begin{tabular}{p{2.5cm}p{2cm}p{2cm}p{2cm}p{2cm}p{2cm}p{2cm}}
  \toprule
   \textbf{}  &  \textbf{Method}  &  \textbf{all bands}  \\ \hline 
   \multirow{4}{*}{\textbf{Transfer Learning}} 
     & TCA \cite{pan2010domain}  & 63.64 / 14.88 \\
     & DANN \cite{ganin2016domain}  & 75.08 / 11.18 \\
     & BiDANN-S \cite{li2018bi}  & 84.14 / 6.87  \\
     & RGNN \cite{zhong2020eeg} & 85.30 / 6.72  \\ \hline
   \multirow{4}{*}{\textbf{Supervised Learning}}
     & SVM \cite{suykens1999least} & 56.73 / 16.29 \\
     & DGCNN \cite{song2018eeg} & 79.95 / 9.02  \\
     & RGNN w/o DA \cite{zhong2020eeg} & 81.92 / 9.35 \\
    &  \textbf{PGCN (ours)} & \textbf{84.59} / \textbf{8.68} \\
  \bottomrule

  \end{tabular}
%   \begin{tablenotes}
%   \footnotesize
% %   \item[*] calculate the average accuracy based on the results of two sessions.
%   \end{tablenotes}
  \end{threeparttable}
  \end{table}

\subsection{Experiment on SEED-IV}

\begin{figure}[h]
  \centering
  \includegraphics[scale=0.55]{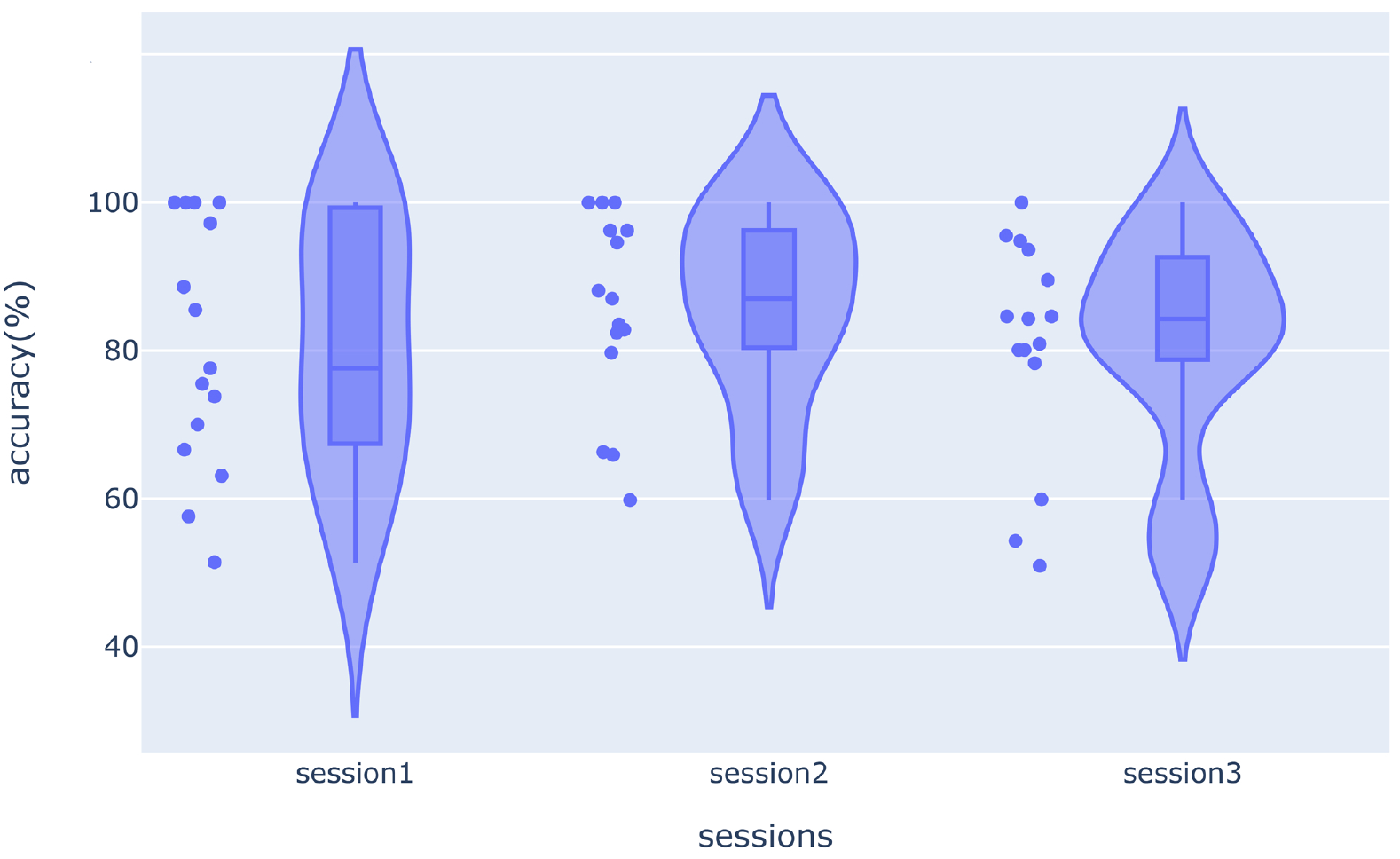}
  \caption{Subject-dependent emotion recognition accuracy on the SEED-IV dataset.}
  \label{figure4}
\end{figure}

Table \ref{table3} shows the results of the proposed PGCN for emotion recognition on SEED-IV. In the subject-dependent experiments, the accuracy of PGCN was 2.87\% higher than that of RGNN under the same experimental setting. In addition, we present the subject-dependent results for different subjects in different sessions in Figure \ref{figure4}. The results show that more than two-thirds of the subjects had an accuracy of more than 80\%, and one-third of the accuracy was above 90\%. In contrast, the emotion recognition results in session 1 were almost evenly distributed between 0.6 and 1, visually demonstrating the considerable variation across subjects under the same experimental setting. 

%%%%%%%%%%%%%%%% table-2
\begin{table}[]
  \centering
  \caption{Subject-dependent and subject-independent classification accuracy (mean/std) on the SEED-IV dataset}
  \label{table3}
  % \begin{tabular}{lllllll}
  \renewcommand{\arraystretch}{1.5}
  \begin{threeparttable}
  \begin{tabular}{m{2.5cm}<{\centering}|m{2.32cm}<{\centering}m{2.52cm}<{\centering}}
  % \begin{tabular}{p{2.5cm}p{2cm}p{2cm}p{2cm}p{2cm}p{2cm}p{2cm}}
  \toprule
   \textbf{Method}  &  \textbf{subject-dependent}  &  \textbf{subject-independent}  \\ \hline 
   SVM \cite{suykens1999least}  & 56.61 / 20.05 & 37.99 / 12.52  \\
   DBN \cite{zheng2015investigating}  & 66.77 / 7.38 & - / -  \\
   DGCNN \cite{song2018eeg} & 69.88 / 16.29 & 52.82 / 9.23  \\
   BiDANN-S \cite{li2018bi} & 70.29 / 12.63 & 65.59 / 10.39  \\
   BiHDM \cite{li2020novel} & 74.35 / 14.09 & 69.03 / 8.66  \\
   RGNN \cite{zhong2020eeg} & 79.37 / 10.54 & 73.84 / 8.02   \\\hline
   BiHDM w/o DA \cite{li2020novel} & 72.22 / 14.69 & 67.47 / 8.22 \\
   RGNN w/o DA \cite{zhong2020eeg} & - / - & 71.65 / 9.43   \\\hline
   \textbf{PGCN (ours)} &  \textbf{82.24}  / \textbf{14.85}  &  \textbf{73.69}  / \textbf{7.16}    \\
  \bottomrule

  \end{tabular}
%     \begin{tablenotes}
%   \footnotesize
%   \item[*] calculate the average accuracy based on the results of two sessions.
%   \end{tablenotes}
  \end{threeparttable}
  \end{table}

% 在subject-independent实验中，相对于除RGNN以外的baselines，PGCN均表现出了超过3%的提升。对于不包含DA模块的RGNN，PGCN也表现出2.04%的提升，而当RGNN加入DA模块后，PGCN也只只稍逊了0.15%。我们发现，BiHDM和RGNN在加入DA模块后分别提升了1.56%和2.19%，表现出领域自适应在降低被试间分布差异时的优异能力。在模型的实际应用中，巧妙的引入DA模块不失为一种提升模型效果的有效手段。
In the subject-independent experiments, the PGCN showed an improvement of over 3\% relative to all baselines except the RGNN. For RGNN without the domain adaption (DA) module, PGCN also showed a 2.04\% improvement, and when the DA module was added to RGNN, PGCN was slightly worse by 0.15\%. We found that BiHDM and RGNN improved by 1.56\% and 2.19\%, respectively, with the addition of the DA module, demonstrating the excellent ability of domain adaptation in reducing distribution diversity between subjects.

\subsection{Experiment on SEED-V}

Table \ref{table4} shows the results of the proposed PGCN for emotion recognition on the SEED-V dataset. All experimental results of baselines are extracted from the corresponding citations. BDAE \cite{zhao2019classification} enhances emotion recognition with the help of high-level representational features extracted by a bimodal deep auto-encoder, MD-AGCN \cite{li2021multi} proposes a multi-domain adaptive graph convolution network that incorporates frequency and temporal domain knowledge, making full use of the complementary information of EEG signals. It can be found that the PGCN improves 0.92\% over the previous best MD-AGCN on the subject-dependent task, while the PGCN is able to achieve 61.78\% accuracy in emotion recognition on the subject-independent task.

  %%%%%%%%%%%%%%%% table-3
\begin{table}[]
  \centering
  \caption{Subject-dependent and subject-independent classification accuracy (mean/std) on the SEED-V dataset}
  \label{table4}
  % \begin{tabular}{lllllll}
  \renewcommand{\arraystretch}{1.5}
  \begin{threeparttable}
  \begin{tabular}{m{2cm}<{\centering}|m{2.52cm}<{\centering}m{2.52cm}<{\centering}}
  % \begin{tabular}{p{2.5cm}p{2cm}p{2cm}p{2cm}p{2cm}p{2cm}p{2cm}}
  \toprule
   \textbf{Method}  &  \textbf{subject-dependent}  &  \textbf{subject-independent}  \\ \hline 
   SVM \cite{zheng2015investigating}  & 69.5 / 10.28 & - / -  \\
   BDAE \cite{zhao2019classification}  & 79.7 / 4.76 & - / -  \\
   MD-AGCN \cite{li2021multi} & 80.77 / 6.61 & - / -  \\\hline
   \textbf{PGCN (ours)} &  \textbf{81.69}  / \textbf{10.57}  &  \textbf{61.78} / \textbf{8.59}    \\
  \bottomrule

  \end{tabular}
%   \begin{tablenotes}
%   \footnotesize
% %   \item[*] calculate the average accuracy based on the results of two sessions.
%   \end{tablenotes}
  \end{threeparttable}
  \end{table}

% baseline： 简单的两层GCN，0/1 adj，没有可学习adj，没有短接
% local：3D，可学习adj，短接
% meso：
% global

\section{Discussion}\label{Discussion}

% 在本章，我们将借助消融实验证明PGCN的各个模块的作用并尝试解读PGCN能够工作良好的原因，并对PGCN的模型结果进行可视化展示。

In this chapter, we demonstrate the role of the various modules of the PGCN with the help of ablation experiments, try to decipher why the PGCN works well. We also visualize the results of the PGCN.

\subsection{Ablation Study}

% 我们对PGCN中的local, mesoscopic和global 三个模块进行了拆解和组合，展示每个模块对于情绪识别的效果，并将结果展示在表5中。在消融实验中，我们只对模型中的local，mesoscopic和global 3个模块所组成的特征提取网络进行修改，并不对图2所示的情绪识别模块及其它代码做任何改动。 在表5中，为了更客观的展示local模块对于网络效果的提升，backbone单独使用时仅表示最基础也最常用的基于电极的2D邻接关系的两层GCN网络，backbone与local同时存在时对应了图2中的local feature aggregation模块。
We disassembled and combined the local, mesoscopic and global modules in the PGCN to demonstrate the effect of each module on emotion recognition and show the results in Table \ref{table5}. In the ablation experiment, we only modified the feature extraction network composed of the local, mesoscopic, and global modules without changing other parts. In Table \ref{table5}, the backbone represents the most basic and commonly used two-layer GCN network based on 2D electrode adjacency matrix \cite{ding2021lggnet, song2021variational}, and backbone module and local module do not activate at the same time.

\begin{table*}[]
  \centering
  \caption{Ablation study for subject-dependent classification accuracy (mean/std) on the SEED and SEED-IV dataset.}
  \label{table5}
  % \begin{tabular}{lllllll}
  \renewcommand{\arraystretch}{1.5}
  \begin{threeparttable}
  \begin{tabular}{m{2.0cm}<{\centering}|m{1.9cm}<{\centering}m{1.7cm}<{\centering}m{2cm}<{\centering}m{1.9cm}<{\centering}|m{2cm}<{\centering}m{2cm}<{\centering}}
  % \begin{tabular}{p{2.5cm}p{2cm}p{2cm}p{2cm}p{2cm}p{2cm}p{2cm}}
  \toprule
  \textbf{Method} &  \textbf{Backbone}  &  \textbf{Local}  &  \textbf{Mesoscopic} &  \textbf{Global}     &  \textbf{SEED}    & \textbf{SEED-IV}     \\ \hline 
   {Baseline} &$\bullet$  & $\circ$  & $\circ$  & $\circ$    & 
   {92.34} /  {7.68}  &  {75.94} /  {13.02}   \\
   
   {Global-only} &$\bullet$  & $\circ$  & $\circ$  & $\bullet$    & 
   {92.93} /  {8.35}  &   {76.36} /  {14.17}  \\
   
   {Local-only} &$\circ$  & $\bullet$  &$\circ$   & $\circ$    & 
   {93.69} /  {6.12}  &  {77.84} /  {14.68}  \\
   
   {Meso-only} &$\bullet$  & $\circ$  & $\bullet$  & $\circ$    & 
   {94.91} /  {6.54}  &  {80.71} /  {12.72}   \\
   
   {Meso-removed} &$\circ$  & $\bullet$  & $\circ$  & $\bullet$    & 
   {94.08} /  {7.16}  &   {77.84} /  {13.08}  \\
   
   {Global-removed} &$\circ$  & $\bullet$  & $\bullet$  & $\circ$    & 
   {95.41} /  {6.71}  &  {80.6} /  {13.27}  \\
   
   {Local-removed} &$\bullet$   & $\circ$ & $\bullet$  & $\bullet$    &
   {96.41} /  {4.54}  &  {80.96} /  {14.52} \\ \hline
   
   \textbf{PGCN}&$\circ$   & $\bullet$ & $\bullet$    & $\bullet$  &  \textbf{96.93} /  \textbf{5.11}  & \textbf{82.24} / \textbf{14.85} \\

  \bottomrule
    \end{tabular}
    \begin{tablenotes}
  \footnotesize
  \item[] In each line, $\bullet$ means the module is employed, and $\circ$ means that module is blocked. The backbone module and local module do not activate at the same time.
  \end{tablenotes}
  \end{threeparttable}

\end{table*}

% 通过添加3个模块，模型在SEED数据集上情绪识别准确率从92.34%提升到96.93%，在SEED-IV数据集上从75.94%提升到82.24%。且可以发现，随着模块的加入，模型的效果呈现总体上提升的趋势。
The three modules improve the sentiment recognition accuracy of PGCN from 92.34\% to 96.93\% on the SEED dataset; and from 75.94\% to 82.24\% on the SEED-IV dataset; furthermore, it seems each module contributes positively. The following is a more detailed discussion of the data in Table \ref{table5}.

% 3个模块各自有一定程度的提升
\begin{itemize}
\item[(1)]The introduction of each module individually gives a comprehensive boost to the model, with the local module having a Sharpley value of about 2\% on the SEED and SEED-IV dataset, the meso module having a Sharpley value of about 3\% - 3.5\% on the SEED and SEED-IV dataset, and the global module having a Sharpley value of about 1\% and 2\% on the SEED and SEED-IV dataset.

% 对比baseline与meso-only模块，meso模块可以对网络拟合能力带来巨大的提升，在seed数据集上提高了2.57%，在seed-IV数据集上提高了4.77%。我们推测该进步可能来自meso 模块能够借助先验知识提取到mesoscopic尺度的节点之间有判别力的feature，并将提取到的特征与GCN提取的特征进行融合，有效的提升图网络的效果。
\item[(2)]Comparing the baseline and meso-only modules, the meso module can effectively improve the network performance, with a 2.57\% improvement on the SEED dataset and a 4.77\% improvement on the SEED-IV dataset. We speculate that the improvement may come from that the meso module can extract discriminative features between nodes at the mesoscopic scale with the reference of a priori knowledge and effectively improve the network by fusing the extracted features with the output of the backbone.

\item[(3)]Since the meso-removed is a combination of the local and global module, comparing it with the local-only module reveals that introducing the global module on top of the local module only improves the emotion recognition ability by 0.39\% on SEED, while it brings no improvement on SEED-IV. We hypothesized that for the GCN-based model, stacking network layers is accompanied by a severe over-smoothing problem, and to verify the conjecture; we conducted a more in-depth experiment in Figure \ref{figure5}. As a comparison, global-only brings about 0.5\% improvement on baseline probably because the GCN in the baseline is not sufficiently trained and optimized.

\end{itemize}

% 由于meso模块包含脑区尺度和左右半脑尺度这两个不同尺度，我们对meso模块中的子模块进行了消融，并将其展示在表6中。

Since the meso module contains two different scales, the brain region scale, and the hemisphere scale, we ablated the submodules in the meso module and presented the result in Table \ref{table6}.

Before the introduction of the meso module, the stacking of local and global modules allowed the network to have the corresponding local and global perceptions but was accompanied by stagnation or even a decrease in the fitting ability due to the deepening of the GCN. By introducing the 7-region and 2-region meso module, the network gains mesoscopic perceptual fields while adding some critical virtual mesoscopic centers, resulting in an improved fitting ability.

\begin{table}[]
  \centering
  \caption{Effectiveness of Meso-layer on the SEED and SEED-IV dataset.}
  \label{table6}
    \renewcommand{\arraystretch}{1.5}
    \begin{threeparttable}
    \begin{tabular}{m{1.3cm}<{\centering}m{1.3cm}<{\centering}|m{1.82cm}<{\centering}m{1.82cm}<{\centering}}
    % \begin{tabular}{p{2.5cm}p{2cm}p{2cm}p{2cm}p{2cm}p{2cm}p{2cm}}
    \toprule
       \textbf{7 regions} & \textbf{2 regions} &  \textbf{SEED}  &  \textbf{SEED-IV}  \\ \hline 
        $\circ$  &$\circ$ & {94.08} / {7.16} & {77.84} / {13.08}  \\
       $\bullet$  &$\circ$ & {94.74} / {6.66}   &  {80.34} / {14.71}  \\
       $\circ$  &$\bullet$ & {94.85} / {5.91}  &  {79.77} / {13.16}   \\\hline
       $\bullet$  &$\bullet$ &  \textbf{96.93} /  \textbf{5.11}  &  \textbf{82.24} / \textbf{14.85}  \\
    \bottomrule
    \end{tabular}
    \begin{tablenotes}
  \footnotesize
  \item[] In each line, $\bullet$ means the module is employed, and $\circ$ means that module is blocked.
  \end{tablenotes}
  \end{threeparttable}
\end{table}

% 在表5和表6中，所使用的超参数配置是针对PGCN优化的，但是对于其它模型却可能存在更优的选择。但是为了控制变量，同时为了减少调参的压力，我们尽力将超参数保持一致，但这也可能使得消融实验所展示的提升比实际的提升可能更大。
In Table \ref{table5} and \ref{table6}, the hyperparameter is optimized for PGCN, but more optimal choices may exist for other models. However, to control the variables and reduce the parameter tuning, we tried to keep the hyperparameters fixed, but this may allow the ablation experiments to demonstrate more significant improvements than model-by-model optimization.

\subsection{Why PGCN works? Network Architecture Analysis}

\begin{figure}[h]
  \centering
  \includegraphics[scale=0.33]{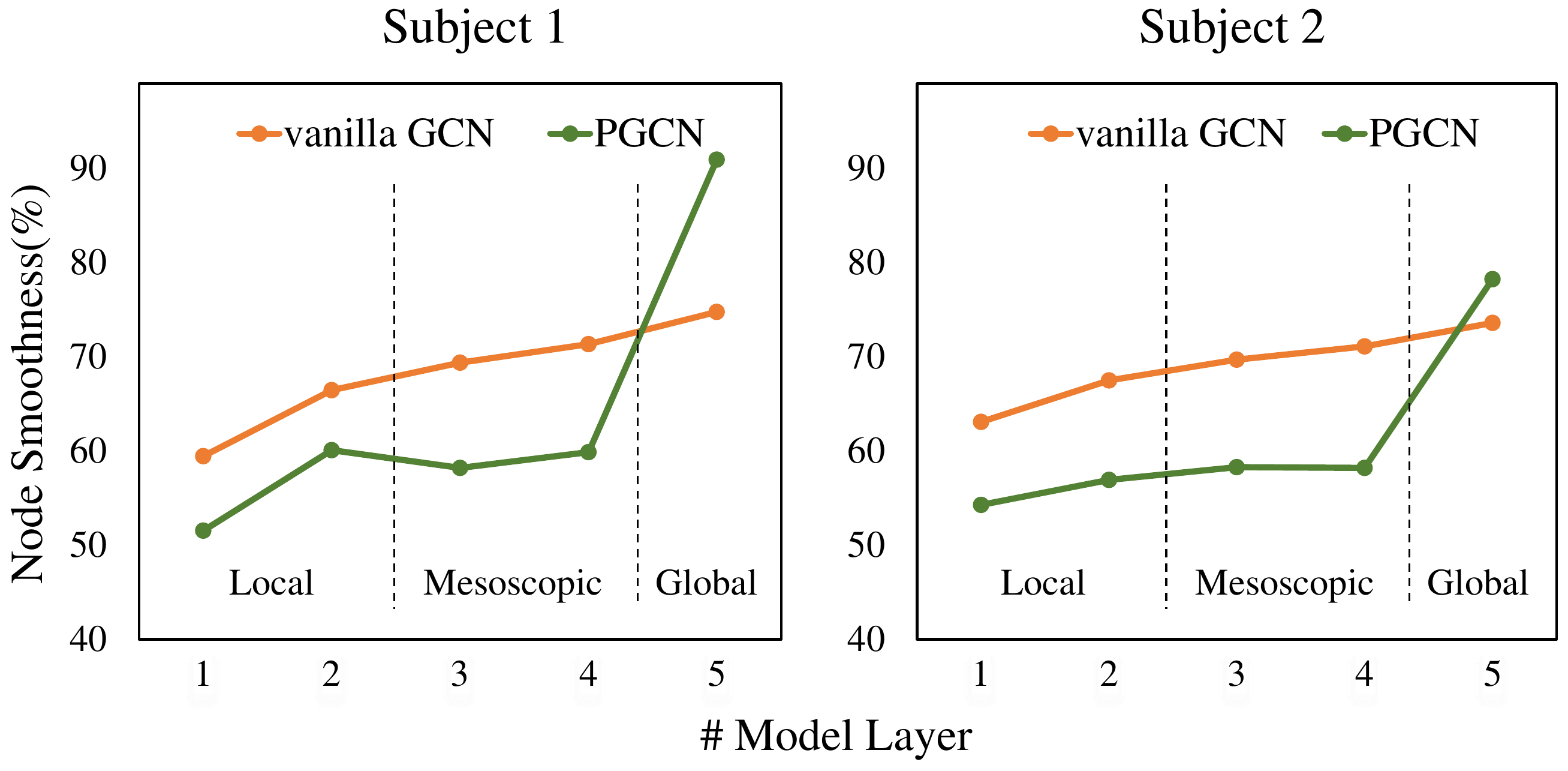}
  \caption{The curve of node smoothness with the increasing number of network layers. Node smoothness calculates the mean of the cosine similarity between the nodes of the output features of each layer.}
  \label{figure5}
\end{figure}

\begin{figure}[h]
  \centering
  \includegraphics[scale=0.32]{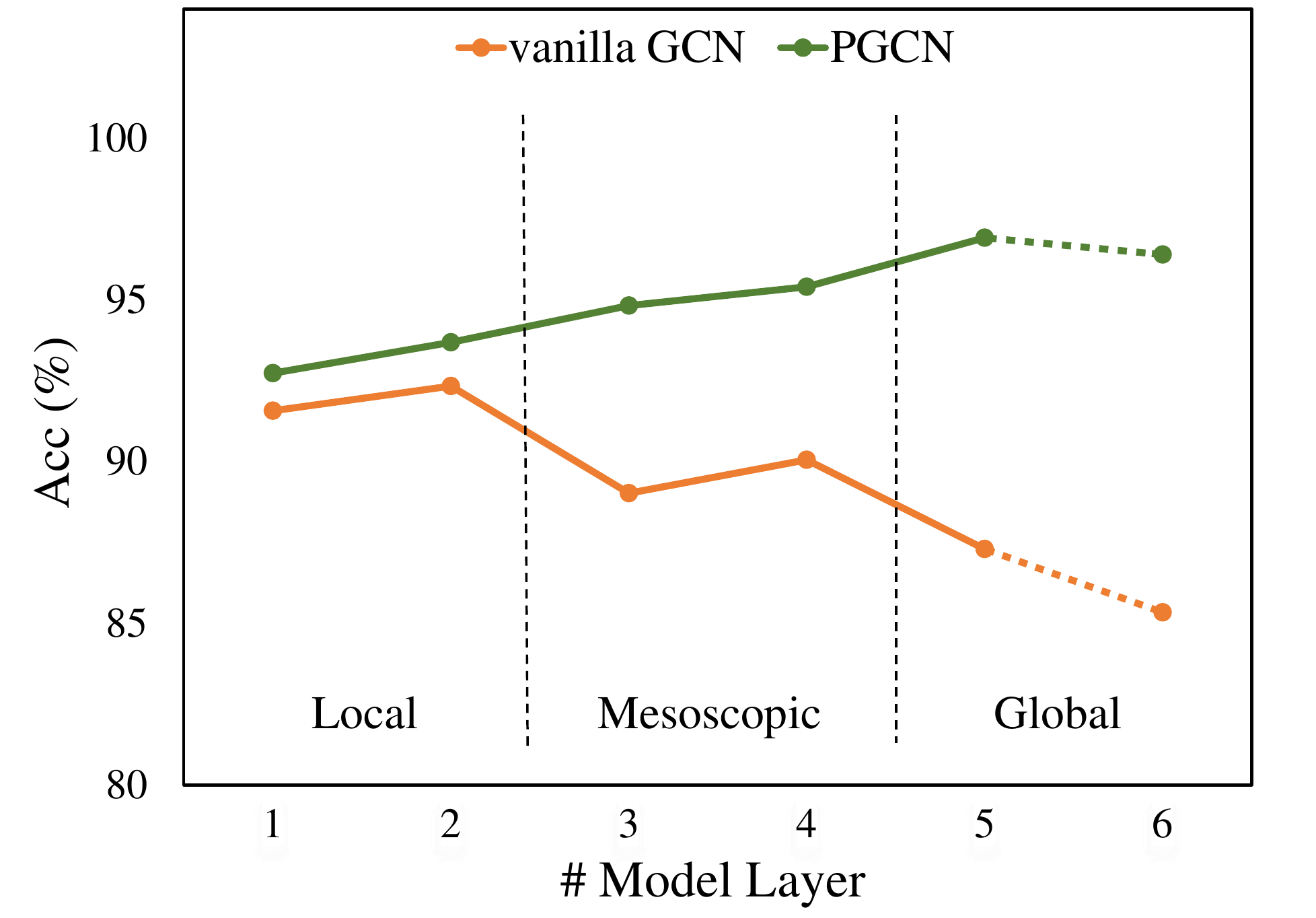}
  \caption{Average prediction accuracy of vanilla GCN and PGCN on the SEED dataset. As the perceptual field increases, the accuracy of the vanilla GCN begins to decline after the second layer, and the fitting ability of the PGCN continues to improve until the fifth layer, when it begins to decline.}
  \label{figure6}
\end{figure}

To further explore why PGCN works, we plot line graphs as the layer number of the GCN network increases and depict it in Figure \ref{figure5} and Figure \ref{figure6}. In this case, vanilla GCN presents the most commonly used GCN network based on a 2D electrode adjacency matrix, and for the 6-layer network of PGCN, we added a attention-based global-scale layer.

We plot the node smoothness curve with an increasing network layer in Figure \ref{figure5} to visualize the over-smoothing problem. In this case, node smoothness calculates the cosine similarity between the nodes of the output features in each layer. Thanks to the redesigned initialized adjacency matrix, the PGCN has a minor increase in node smoothness than the vanilla GCN at the local layer \cite{chen2020measuring}. At the mesoscopic layer, with the introduction of virtual nodes, the node smoothness remains almost constant or even decreases while the PGCN gains a larger perceptual field, while the node smoothness of the vanilla GCN continues to increase. At the global layer, the global perception field allows the node smoothness of the PGCN to rapidly increase and surpass that of the vanilla GCN.

Figure \ref{figure6} shows the average prediction accuracy of the networks with different layer numbers. For vanilla GCN networks, as the layer number increases, the recognition accuracy tends to decrease after reaching a maximum of 92.34\% at two layers, which is why the majority of current GCNs for emotion recognition tasks have two-layer networks \cite{ding2021lggnet, song2021variational}. In the PGCN, as the number of network layers increases, the recognition accuracy reaches a maximum of 96.93\% at the number of layers of five and begins to decline.

% 对比basic GCN 与 PGCN 可以发现，得益于引入了稀疏的图关系推理和局域特征聚合，PGCN得以在浅层网络上（layer<=2）领先basic GCN 1.2%。当网络继续加深后（2<layer<=4），basic      GCN中更大感受野带来的更好的情绪识别效果在与over-smoothing带来的特征相似性过高问题的竞争中逐渐处于劣势，模型的拟合能力开始下降；在PGCN中，得益于网络能够以一种高度稀疏的方式聚合先验知识指导的mesoscopic尺度的特征，特征被聚合到虚拟节点，在得到了更大的感受野的同时并未为电极节点带来特征聚合而导致的特征相似性过高问题，PGCN模型的拟合能力得以继续增加。当网络达到全局感受野后（layer>4）,basic GCN 中的over-smoothing问题更加凸显，使得模型的拟合能力进一步下降；在PGCN中，通过将电极节点与虚拟节点进行融合，并使用稀疏化的基于注意力的邻接矩阵，在网络层数为5时，更大感受野带来的更好的信息交互在与over-smoothing带来的特征信息过度交互的竞争中依旧能保持优势，模型的拟合能力继续提升，但是随着网络层数继续增加，over-smoothing问题逐渐显著，模型的拟合能力开始下降。
Comparing vanilla GCN with PGCN, it can be seen that thanks to the introduction of sparse graph relational reasoning and local feature aggregation, PGCN can outperform vanilla GCN by 1.2\% on shallow networks (layer $\leq$ 2). When the network continues to deepen (2 $<$ layer $\leq$ 4), the better emotion recognition effect brought by the larger receptive field in vanilla GCN is gradually at a disadvantage in the competition with the high feature similarity problem caused by over-smoothing, the effectiveness begins to decline; in the PGCN, thanks to the network's ability to aggregate mesoscopic features guided by a priori knowledge in a highly sparse manner, the features are aggregated to virtual nodes, which gives a larger perceptual field without causing excessive feature similarity problems for the electrode nodes, and the network performance continues to increase. The over-smoothing problem in the vanilla GCN becomes more pronounced when the network reaches the global perceptual field (layer $>$ 4), which further reduces the model's effectiveness; in the PGCN, by fusing electrode nodes with the virtual nodes and employing a sparse attention-based adjacency matrix, the better information interaction due to increasing the perceptual field still competes with the over-interaction of feature information due to over-smoothing at network layer number 5, and the model's performance continues to improve, but as the network layers continue to stack, over-smoothing prevails, and the model's performance declines.

% GCN is excellent at aggregating emotion-related EEG signals but with an equally significant over-smoothing problem. PGCN provides better node signal interaction at different scales by stacking pyramidal networks and suppresses the over-smoothing problem by introducing a priori information to guide the virtual key nodes and sparse relationships to obtain better emotion recognition ability.

\subsection{Representation Visualization}

\begin{figure*}
	\centering
	\subfigure(a){
	\includegraphics[width=0.95\textwidth]{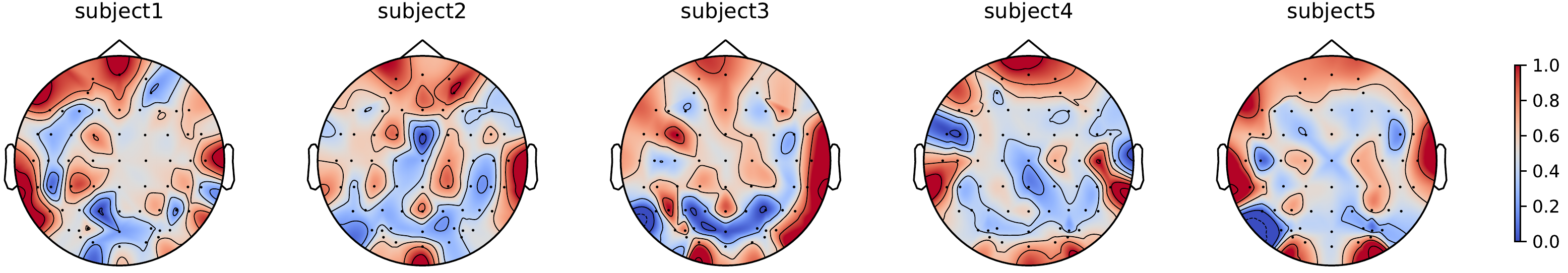}} \\ 
	\subfigure(b){
	\includegraphics[width=0.95\textwidth]{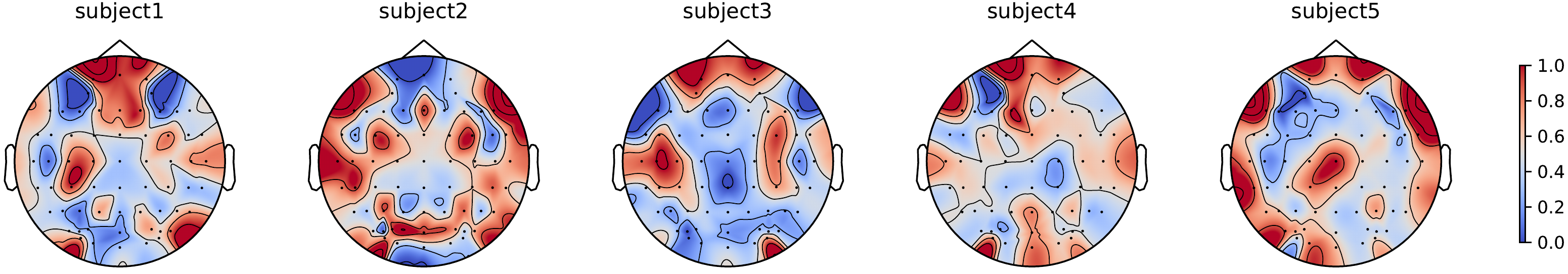}}
	\caption{Heat map of the learned adjacency matrix diagonal elements for subject-dependent emotion recognition on the (a) SEED and (b) SEED-IV datasets, the diagonal elements were deflated to between 0 and 1 for presentation. For most subjects, electrodes in the frontal, temporal and occipital lobes have greater self-weighting.}
	\label{figure7}
\end{figure*}

To manifest brain activation in the emotion recognition progress, we selected five subjects in each of the SEED and SEED-IV datasets and displayed their heat map of the learned adjacency matrix diagonal elements in Figure \ref{figure7}. The diagonal elements of the adjacency matrix provide the most intuitive indication of the weights of the node features in the graph convolution, and the diagonal elements of the adjacency matrix were deflated to between 0 and 1 for presentation. 

% 结合两个数据集的结果，可以发现，位于额叶、颞叶和枕叶的电极在经过学习后在图卷积中获得了更大的权重。其中，在3分类的SEED数据集上，位于颞叶和枕叶的节点有明显的激活，位于额叶的节点部分激活。在4分类的SEED-IV数据集上，不仅颞叶和枕叶的节点有明显的激活，位于额叶的节点激活更加显著。我们猜测，导致两个数据集上的节点的激活差异可能是由于对于3分类和4分类任务存在一定程度的归纳偏置差异，其中，4分类任务需要同时关注额叶、颞叶和枕叶的信息才能实现准确的情绪识别。
Combining the results from the two datasets, electrodes in the frontal, temporal and occipital lobes gained greater weight after learning. In particular, on the 3-category SEED dataset, there was significant activation of nodes located in the temporal and occipital lobes and partial activation of nodes located in the frontal lobe. On the 4-category SEED-IV dataset, nodes were significantly activated in the temporal and occipital lobes and more significant activation in the frontal lobes. The difference in activation of the nodes on the two datasets may be caused by a degree of inductive bias for the 3-category and 4-category classification tasks, where the four classification task requires attention to information from the frontal, temporal and occipital lobes simultaneously for accurate emotion recognition.

\begin{figure}
	\centering
    \subfigure(a){
	  \includegraphics[width=0.44\linewidth]{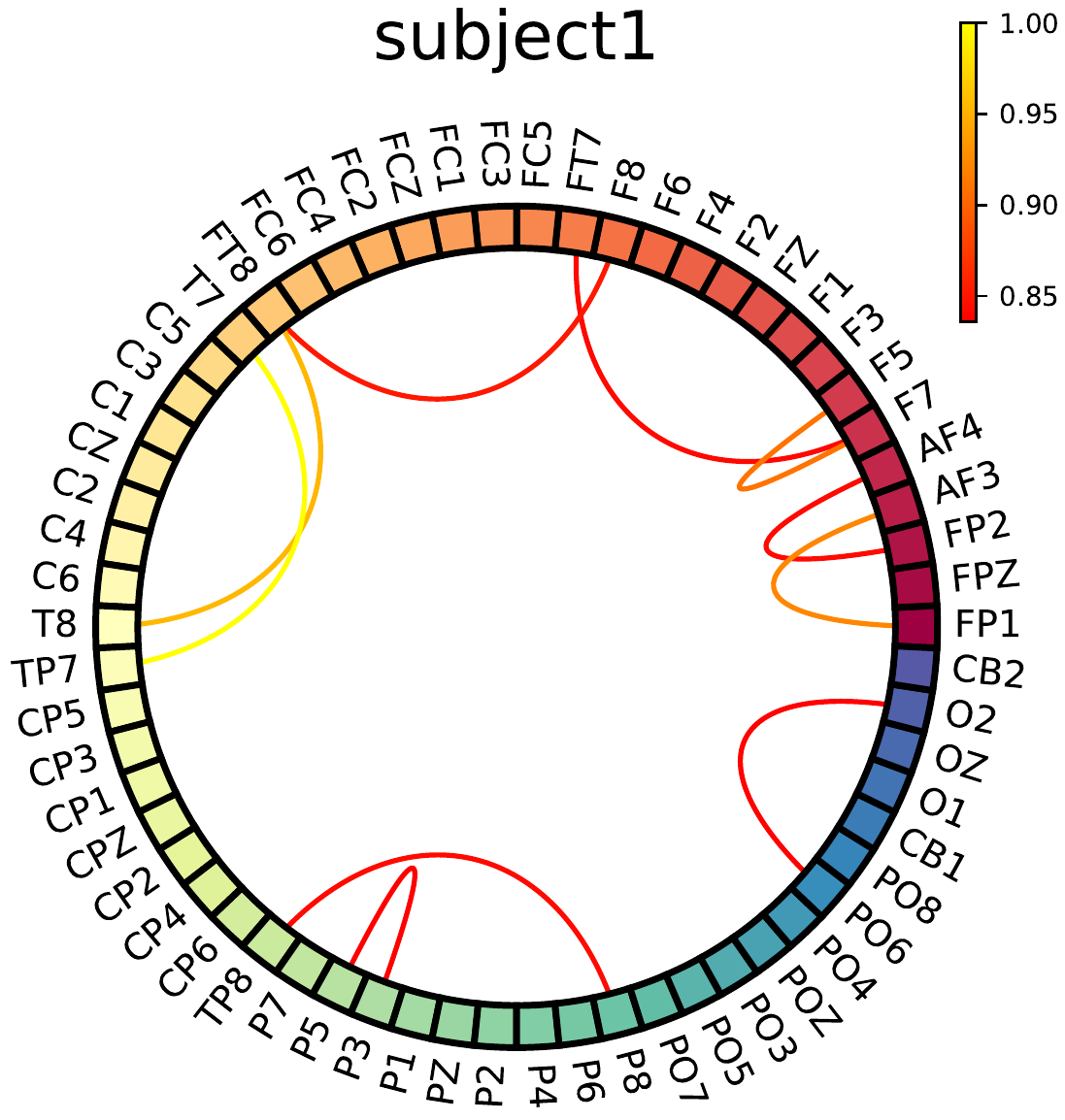}\label{}
      \includegraphics[width=0.44\linewidth]{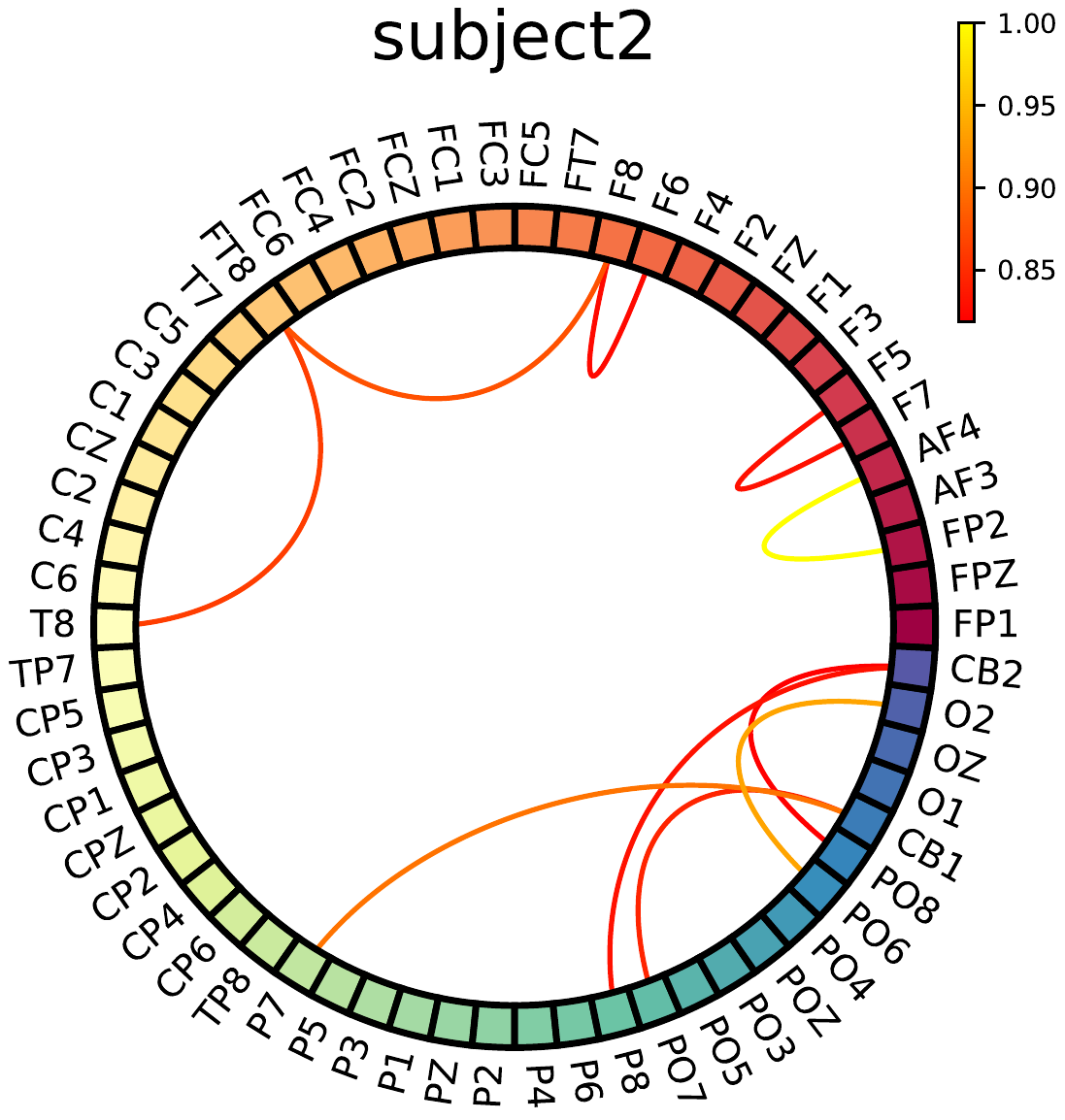}\label{}} \\
    \subfigure(b){
      \includegraphics[width=0.44\linewidth]{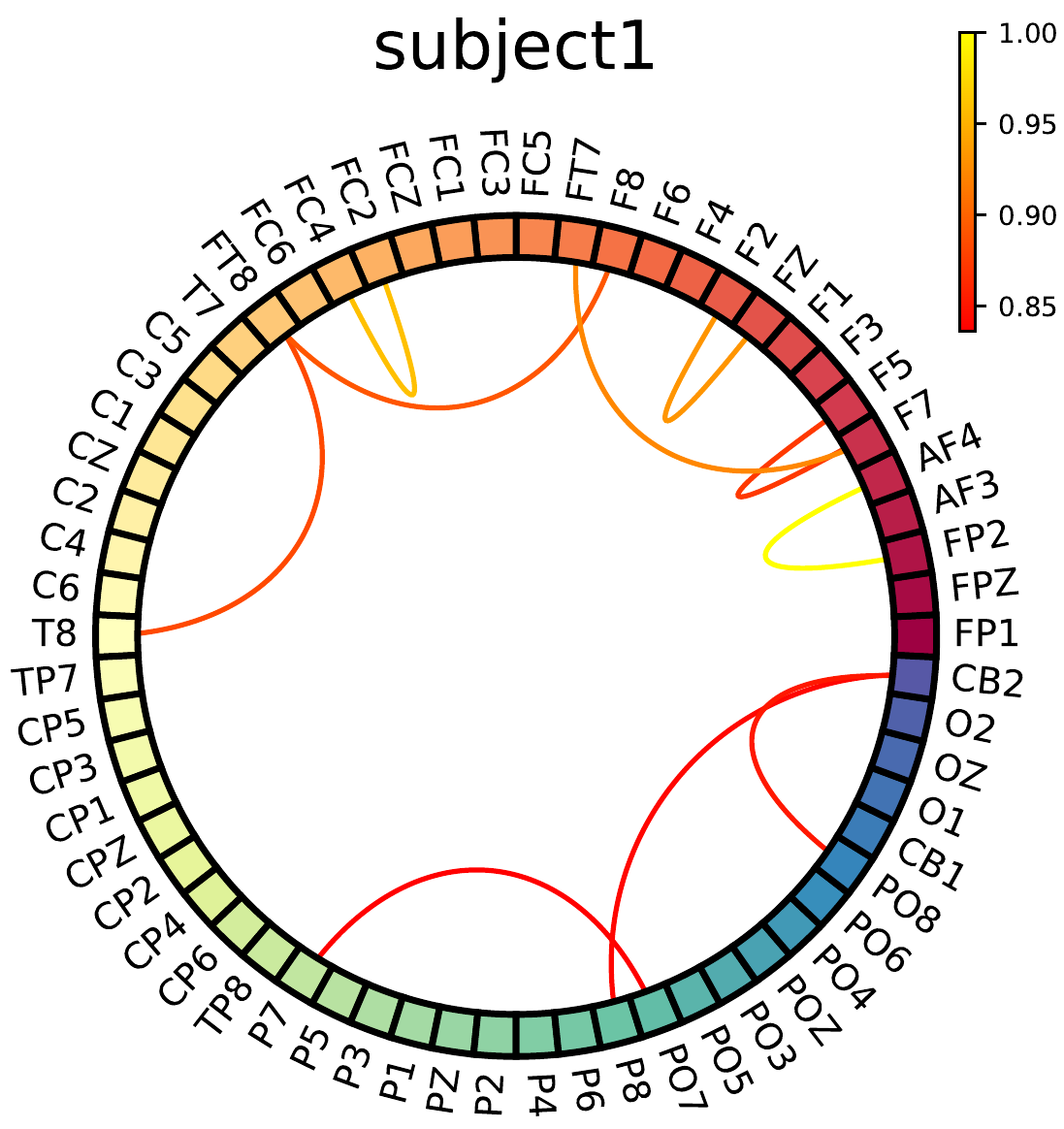}\label{}
      \includegraphics[width=0.44\linewidth]{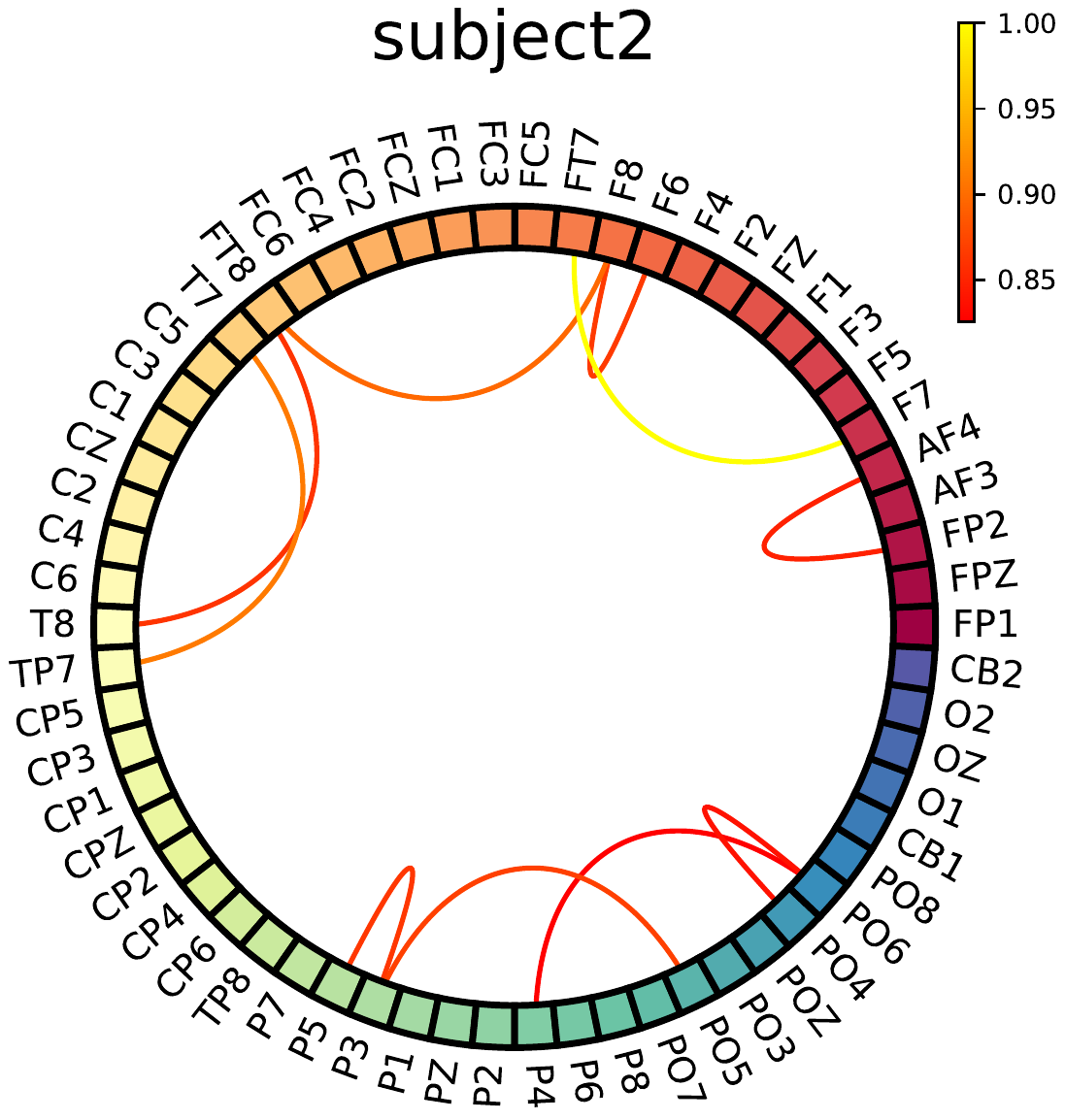}\label{}}
	\caption{Top 10 connections of learned adjacency matrix on the (a) SEED dataset and (b) SEED-IV dataset, the electrodes occupying the most important connections are distributed in the frontal, temporal and occipital lobes.}
	\label{figure8}
\end{figure}

% 为了展示不同的节点之间的连接关系，我们在SEED数据集和SEED-IV数据集各选取了两个被试，绘制不同被试的学习后的邻接矩阵的top-10连接。与heat map的绘制不同，我们去除了对角线上的元素，只绘制了不同节点间的连接。
To show the connections between nodes, we selected two subjects in each of the SEED and SEED-IV datasets, and plot the top-10 connections of the learned adjacency matrix in Figure \ref{figure8}. Unlike the heat map plotting, we removed the diagonal elements and plotted the connections.

Comparing the different chordograms in Figure \ref{figure8}, the top-10 connections vary considerably across subjects. For example, for subject1 of the SEED dataset, the primary connections are mainly in the frontal and temporal lobes, with fewer links in the occipital lobe; in contrast, for subject2, the critical connections are located in the temporal and occipital lobes, while the frontal lobe decreases in importance. For the SEED-IV dataset, a similar situation exists. Throughout the two datasets, although there were some variations in critical connectivity between subjects, most critical connections were consistently related to the temporal, frontal, and occipital lobes, which are closely related to emotion and vision when combined with the chord diagrams of the two data sets.

% 结合图6和图7，尽管存在跨被试和跨数据集的被试相关的差异性，但是对于情绪识别任务，位于额叶、颞叶和枕叶的节点的特征具备更大的自环权重，不仅如此，通过构建与其它节点的强连接，位于上述脑区的节点得以将自身的特征在网络中更多的保留下来，最终实现更准确的情绪识别。
In combination with Figures \ref{figure7} and \ref{figure8}, despite the cross-subject and cross-dataset subject-related variability, the features of nodes located in the frontal, temporal and occipital lobes have greater self-loop weights, and not only that but by building strong connections with other nodes, nodes located in the regions can play a leading role in the network, ultimately achieving excellent emotion recognition.

\section{Conclusion}\label{Conclusion}

% 在本文中，我们提出了一种针对与情绪识别任务的金字塔型图卷积网络。该网络通过构建一个视野渐进增加的图卷积网络，有效的在图卷积的扩大感受野和随之而来的over-smoothing问题中找到平衡，从而有效的提升了情绪识别的准确率。本文分别设计了三个模块获取不同尺度的电极之间的信息，其中，local模块主要学习邻近节点之间的强烈的局部偏置（局域特征），meso模块借助先验知识学习脑区尺度的节点之间的联系，global模块则学习跨区域的全局节点之前的稀疏的联系，并将3个模型学习到特征进行有效的整合，从而提升了情绪识别的效果。实验证明提出的PGCN在三个公开数据集上都能取得最佳的结果。我们之后的工作将会聚焦于：1）设计优化的meso结构，在获取到有鉴别力的脑区特征的同时，进一步的降低over-smoothing的效果；2）探索更有效的面向EEG情绪识别任务的脑区邻接矩阵构造，获取到更有效的初始邻接矩阵；3）尝试引入更多的脑研究的先验知识到情绪识别任务中，提升情绪识别的准确率。
This paper proposes a pyramidal graph convolution network that improves emotion recognition accuracy by effectively finding a balance between the expanded receptive fields of GCN and the consequent over-smoothing problem. In PGCN, three separate modules are designed to acquire information between electrodes at different scales. The local module mainly concentrates on small-world properties, the mesoscopic module learns connections between different brain regions constructed with priori knowledge, and the global module learns the sparse connections between global nodes. The features learned by the three modules are then effectively integrated to improve the effectiveness of emotion recognition. The proposed PGCN achieves the best results on all three publicly available datasets. Our future work will focus on 1) designing optimized meso structures to reduce over-smoothing further while acquiring discriminatory features; 2) exploring a more effective initial brain adjacency matrix; 3) trying to introduce more priori knowledge to improve the emotion recognition accuracy.

\appendices
% \section{fd}

% \section*{Acknowledgment}

% This work was supported in part by National Natural Science Foundation of China (62106248), Zhejiang Provincial Natural Science Foundation of China (LQ20F030013), Ningbo Public Service Technology Foundation, China (202002N3181), and Medical Scientific Research Foundation of Zhejiang Province, China (2021431314).

\ifCLASSOPTIONcaptionsoff
  \newpage
\fi

\bibliographystyle{IEEEtran}
% \bibliographystyle{plainnat}

% \normalem

\bibliography{bare_jrnl}

% H.~Kopka and P.~W. Daly, \emph{A Guide to \LaTeX}, 3rd~ed.\hskip 1em pdlus

% \end{thebibliography}


\newpage

\section{} \label{model}

This paper uses PyTorch to build PGCN and deploy it on a 2080TI GPU. The local module contains two layers of GCN networks with a size of 62 $\times$ 5 for the input and 62 $\times$ 30 for the output; the mesoscopic module contains two layers of networks with an output size of 71 $\times$ 30; the Global network contains one layer of GCNs with an output size of 71 $\times$ 70; the emotion recognition network contains 
a 3-layer fully connected network. For the optimization of the model, we used the AdamW optimizer and warm-up, setting the learning rate to 1e-2 and the batch size to 64. We evaluated the model using the average accuracy (ACC) and standard deviation (STD).

\section{}\label{Dataset}

\subsubsection{SEED Dataset}

The SEED dataset collected EEG emotional data from 15 subjects (seven males and eight females) who watched movie clips with three different emotional tendencies: negative, positive, and neutral. Three sessions of EEG data were collected from each subject, each containing 15 movie clips of different emotions. Each movie clip corresponds to a trial, and each trial contained a 5-second hint, a 4-minute movie clip, a 45-second self-assessment, and a 15-second break. 

In the subject-dependent experiments, we followed the experimental setup of \cite{zheng2015investigating, song2018eeg, li2018bi}, for each subject, we used the first nine trials of the same session as the training set and the last six trials as the testing set, and calculated the mean and standard deviation for all subjects on two sessions \cite{zheng2015investigating, zhong2020eeg, song2021variational}. In the subject-independent experiments, we followed the experimental setup in \cite{song2018eeg, li2018bi} and implemented a leave-one-subject-out cross-validation, and calculated the mean and standard deviation of all subjects on all three sessions.

\subsubsection{SEED-IV Dataset}

The SEED-IV dataset collected EEG emotional data from 15 subjects (7 males and eight females) who watched movie clips with four emotional tendencies: neutral, sad, fear, and happy. Each subject watched 24 movie clips of three sessions, and each movie clip corresponding to a trial, and each trial lasted around 2 minutes.

In the subject-dependent experiments, we followed the experimental setup of \cite{li2020novel, zhong2020eeg}, and for each subject, to ensure data balance, we first set aside the last two trials of each emotion data in each session as the testing set and use the remaining 16 trials as the training set. In the subject-independent experiments, we followed the experimental setup of \cite{li2020novel, zhong2020eeg} and implemented cross-validation with one subject left behind. In evaluating the results, we calculated the average accuracy of all subjects in the three sessions.

\subsubsection{SEED-V Dataset}

The SEED-V dataset collected EEG data from 16 subjects (six males and ten females) and contained five emotional tendencies: happy, disgust, neutral, fear, and sad. Each subject watched 15 movie clips in three sessions.

% 在被试相关的实验中，遵循之前的实验设定，我们将每一个被试的每一个session的前10个trials的数据作为训练集，将后5个trials作为测试集，并计算每一个被试在每一个session上的情绪识别的效果，最终计算出在所有被试在3个session上情绪识别的均值和标准差。对于被试无关的实验，我们同样遵循了留一的实验方法，分别将同一个session中不同的被试轮流作为测试集，余下的被试作为训练集，并计算出所有被试在3个session上的情绪识别的准确率和标准偏差。
In the subject-dependent experiments, following the previous experimental settings \cite{zhao2019classification,li2021multi}, we used the data of the first ten trials as the training set, and the last five trials as the testing set. For the subject-independent experiments, we also followed the same leave-one-out experimental approach. We calculated the accuracy and standard deviation of all subjects on the three sessions on both two experiments.




%\vfill


\end{document}